\documentclass[a4paper,fleqn]{cas-sc}

\usepackage[authoryear]{natbib}

\usepackage{times}
\usepackage{latexsym}
\usepackage{subcaption}

\usepackage{url}

\usepackage{natbib}
\usepackage{algorithm}
\usepackage{algorithmic}
\usepackage{listings}
\usepackage{graphicx}
\usepackage{subcaption}
\usepackage{color}
\usepackage{multirow}
\usepackage{array}
\usepackage[shortcuts]{extdash}
\usepackage[inline]{enumitem}
\usepackage{soul}
\usepackage{cancel}
\usepackage[section]{placeins}
\graphicspath{ {images/} }

\begin{document}
\shorttitle{Experiments in Extractive Summarization}
\shortauthors{Daniel Lee et~al.}

\title [mode = title]{Experiments in Extractive Summarization: Integer Linear Programming, Term/Sentence Scoring, and Title-driven Models}
%%%%%%%%%%
% TITLE NOTE MARKS
% \tnotemark[1,2]
%
% \tnotetext[1]{This document is the results of the research
%   project funded by the National Science Foundation.}
%
% \tnotetext[2]{The second title footnote which is a longer text matter
%   to fill through the whole text width and overflow into
%   another line in the footnotes area of the first page.}
%%%%%%%%%

\author[1]{Daniel Lee}[orcid=0000-0001-6393-3105]
\cormark[2]
\ead{dljr0122@cs.uh.edu}

\author[1]{Rakesh Verma}
\cormark[1]
\ead{rmverma@cs.uh.edu}

\author[1]{Avisha Das}
\ead{adas5@uh.edu}

\author[1]{Arjun Mukherjee}
\ead{arjun@cs.uh.edu}

% \author[1]{Uroosa Ali}
% \ead{uroosa.234@gmail.com}
% 
\address[1]{University of Houston, TX, USA}

\cortext[cor1]{Corresponding author}
\cortext[cor2]{Principal corresponding author}

\begin{abstract}
In this paper, we revisit the challenging problem of unsupervised  single-document summarization and study the following aspects:
\begin{enumerate*}[label={(\roman*)}]
	\item Integer linear programming (ILP) based algorithms,
    \item Parameterized normalization of term and sentence scores, and
    \item Title-driven approaches for summarization.
\end{enumerate*}
We describe a new framework, NewsSumm, that includes many existing and new approaches for summarization including ILP and title-driven approaches. NewsSumm's flexibility allows to combine different algorithms and sentence scoring schemes seamlessly. Our results combining sentence scoring with ILP and normalization are in contrast to previous work on this topic, showing the importance of a broader search for optimal parameters. We also show that the new title-driven reduction idea leads to improvement in performance for both unsupervised and supervised approaches considered. 
\end{abstract}

% \begin{graphicalabstract}
% \includegraphics{figs/grabs.pdf}
% \end{graphicalabstract}

\begin{highlights}
\item We do in depth investigation of parameterizing different scoring schemes.
\item We leverage integer linear programming to extractive summarization.
\item We create new formulations of integer linear programming for summarization.
\item We show our tool, NewsSumm, can be used to efficiently improve supervised methods.
\end{highlights}


\maketitle

%%%%%%%%%%%%%%%%%%%%%%%%
%%%%% INTRODUCTION %%%%%
%%%%%%%%%%%%%%%%%%%%%%%%
\section{Introduction}

Manual summarization of text documents is both expensive and time consuming~\citep{lin2004rouge}. Thus development of automated systems for text summarization is a topic of considerable interest among the research community. Research on automated summarizers began in the late 50s~\citep{luhn1958automatic}, with a system that uses term frequencies for assigning weights to sentences to build a summary.

Summarization research has come a long way since then with automated summarizers aiming to generate the best possible summary. Summaries can be classified into two groups - 
\begin{enumerate*}[label={(\alph*)}]
\item {\em Abstractive}: These are generally created by using lexically similar words and phrases to report the important information in the original article. Such a summary avoids using sentences verbatim from the source. It could also use other vehicles such as word graphs for displaying the important content. 
\item {\em Extractive}: Such a summary is a collection of sentences from the source document purportedly containing the most relevant information. 
\end{enumerate*}

Summaries can be single or multi-document based on the number of documents used as the source material. With the DUC summarization conference, extractive multi-document summarization gained more popularity over single document summarization right after 2001-2002.
A reason for this sudden drop was the absence of a system that could beat the lead-based baseline summaries~\citep{svore2007enhancing} for news articles. 
Thus, finding innovative techniques for single document summarization has a set of challenges different from that of multi-document summarization~\citep{barrera2012combining}.

Our research returns to unsupervised single document summarization. We focus on unsupervised summarization, since: (i) it does not need annotated datasets, and (ii) it can be  orders-of-magnitude faster as neither training nor hyper-parameter optimization phases are necessary. We investigate several new aspects of this topic and make multiple contributions:
\begin{enumerate}
\item We parameterize the most promising word scoring methods to improve ranking of words and also sentence score normalization  (Section~\ref{sec-norm}). By parameterizing the scoring methods, we show that the optimal scores are achieved when parameter values are rarely equal to 1, which is how they are often used in previous literature (Section~\ref{sec:results}).
\item We utilize integer linear programming (ILP) to solve a maximum coverage model of summarization (Section~\ref{sec-ilp}) and extend ILP formulation with budget constraint and with alternative scoring metrics (Section~\ref{sec-budget}). The combination of ILP with $tfidf$ proves to be the best of the unsupervised methods researched (Section~\ref{res:score-ilp}).
\item We leverage titles to drive the modeling of extractive summarization (Section~\ref{sec-title}). In particular, we show that, using our title-driven reduction idea, can also improve even a state-of-the-art  {\em supervised} summarization algorithm as well (Section~\ref{res:title}).
\end{enumerate}

In addition to these contributions, we incorporate the algorithms and approaches into a tool, NewsSumm, that can be used as an extractive summarizer. Since the output comes directly from the original document, the summary can be used as a part of a larger pipeline (e.g. NewsSumm used to extract salient information of target document).  

The rest of this paper is organized as follows. We discuss the most closely related previous work in the next section. In Section~\ref{sec:nsumm}, we introduce NewsSumm, our ILP formulations, parametrizations, and title-driven reduction ideas. In Section~\ref{sec:expt}, we present the experiments and the baselines for comparisons. The quantitative and qualitative results and discussion are in Section~\ref{sec:results} and Section~\ref{sec:concl} concludes. 

%%%%%%%%%%%%%%%%%%%%%%%%
%%%%% RELATED WORK %%%%%
%%%%%%%%%%%%%%%%%%%%%%%%
\section{Related Work}
Summarization of articles like news, research papers, blog posts etc. has always been a difficult task for both humans as well as machines. Automated summarizers have come a long way since the 1950s with current systems using supervised learning methods like the use of submodular optimization models~\citep{sipos2012large} for determining maximum information coverage with minimum redundancy, theme-based summarization using rank-based clustering~\citep{yang2014enhancing}. Unsupervised approaches for text document summarization used in previous research include greedy selection algorithms like Maximal Marginal Relevance~\citep{carbonell1998use} as well as dynamic programming algorithms~\citep{mcdonald2007study}, which view document summarization as a `knapsack problem.' Summarization is a vast topic so we review only the most closely related previous work here, and refer the reader to \citet{gambhirG17,dong18} for more comprehensive overviews. Naturally, we focus on single document summarization. 

\subsection{Extractive Summarization}
%ferreiraCL13
{\bf Scoring Methods.} A proposed syntax and semantic based approach in \citet{barrera2012combining} uses a scoring metric for ranking sentences from the original document, based on the semantics, word popularity as well as sentence position features. This approach outperforms the baseline when tested on a set of scientific magazine articles. This was followed by the use of vertex cover for summarization in \citet{kumarKV13}. \citet{garciaherdandez:09} investigate different weighting schemes based on the terms of a document. A number of word, sentence and graph scoring methods were investigated in \citet{FerreiraCLSFCLSF13}, with tfidf emerging as the winner for word scoring methods. However, the researchers did not parameterize any of the scoring methods as we do in this paper. 
% \textcolor{red}{, but they only consider n-grams with $n > 1$.}

Integer Linear Programming (ILP) has been used to improve efficient summarization. \citet{martinsS09} used ILP to balance both sentence compression and extraction. ILP was studied for concept weighting by \citet{oliveiraLL16}, where sentence position and bigrams were found to be important, and for coherent summarization  by \citet{garciaLE18}, where an entity driven ILP approach was adopted. Extractive summaries can be generated from single documents with the use of genetic operators and guided local search algorithms~\citep{mendoza2014extractive}. 

%\subsection{Clustering}
\noindent {\bf Clustering.} A common approach for unsupervised methods is to use clustering. Then sentences can be selected from different clusters to best represent the document. The K-means clustering algorithm \citep{zhangL09} and Expectation-Maximization algorithm \citep{LedenevaGSRG11} are two approaches that use this approach. Clustering has also been used in multi-document summarization \citep{GuptaS12}.

\noindent {\bf Evaluation of Summaries.} ROUGE, which is based on n-gram overlap of a summary with a gold-standard summary, is a popular method for summary evaluation. Another approach is Pyramid, which involves more human labor. 

\noindent {\bf Limit on ROUGE Recall and F-1 scores}. In ~\citet{Verma:17}, limits on ROUGE recall and F-1 scores are proved, using the idea that no {\em extractive} summarizer can do  better than just returning the entire document as a summary, for DUC datasets for both single-document and multi-document extractive summarization. 

\noindent {\bf Tools.} A number of  tools for automatic summarization have been described in the literature. DocSumm is a Python based extraction based summarizer \citep{Verma:17}. CaseSummarizer is also Python-based but it is built specifically for the domain of legal texts \citep{polsley2016casesummarizer}.  PKUSUMSUM is a general purpose summarizer like DocSumm; however, it is for the Java programming language \citep{zhang2016pkusumsum}.

Recently, SummIT \citep{feigenblat2017summit} is work done by researchers to apply summarization as a direct tool for human users. Humans must input a query to search for documents that they are seeking. However, SummIT's goal is more in application of summarization than improving summarization itself. 
% \textcolor{red}{Unlike NewsSumm none of these tools are publicly available for further development or future work. - This is not true, right? PKUSUMSUM and DocSumm are available, for example.}

There are also plenty of online tools that provide automatic summarization. Doing a simple Google search shows the prevalence of such tools.\footnote{\url{https://www.tools4noobs.com/summarize/}}\footnote{ \url{https://www.textcompactor.com/}} These tools fail to provide an avenue  for further research, e.g., there is no method for bulk/batch summarization. The tools also do not provide a way to use them through scripts or program calls, which further hinders efficient evaluation. NewsSumm lends itself naturally to bulk and unattended testing.

%\subsection{Graph Based Summarization Models}
\noindent {\bf Graph-based Models.} A stochastic graph-based approach LexRank is proposed in~\citet{erkan2004lexrank} for multi-document summarization, which builds the graph based on similarity of sentences, which are the nodes. The summary is generated by using the concept of sentence similarity. In~\citet{mihalcea2004textrank}, TextRank uses the concepts of popular graph based scoring algorithms like PageRank~\citep{page1999pagerank} and HITS~\citep{kleinberg1999web} for sentence similarity scoring.
GRAPHSUM in~\citet{baralis2013graphsum} uses association rules for representing correlations among multiple terms for building a graph based model. 

In~\citet{Verma:17}, researchers frame the summarization problem as a bipartite graph. One set of nodes are the sentences themselves and the other disjoint set being ``thought units.'' The research frames the goal of summarization as a minimal subset of the sentence nodes that covers all the ``thought units.'' The DocSumm tool focuses on a special case of this model that treats the words of the sentence as the ``thought unit''.
 We call this the {\em set-cover} model of summarization. 

\subsection{Abstractive Summarization}
Several methods up to 2017 have been collected by \citet{rachabathuni2017survey}. Recently the research on abstractive summarizers has been predominantly based on neural networks. Some of the flavors of the neural network architectures are encoder/decoder \citep{xu2017decoupling}, attention models \citep{see2017get, CohanDKBKCG18} and GANs \citep{LiuLYQZL18}.

\subsection{Submodular Summarization Models}
The approach of ~\citet{lin2010multi, lin2011class} uses a submodular scoring function, which takes into account the inter-sentence similarity for generating summaries so that there is maximum coverage of important information and redundancy is simultaneously reduced. However, sentence pairwise model in ~\citet{lin2010multi} needs to be manually tuned to get the best inter-sentence similarity measure. The authors of~\citet{sipos2012large} improve  upon this technique and propose a supervised approach to learn both the appropriate similarity measure as well as the required trade-off between coverage and redundancy from the training data. This model was further used in~\citet{sipos2012temporal} for corpus summarization over a time interval.

%%%%%%%%%%%%%%%%%%%%
%%%%% NEWSUMM %%%%%%
%%%%%%%%%%%%%%%%%%%%
\section{NewsSumm}\label{sec:nsumm}
 The NewsSumm tooplkit includes all the previous algorithms available in DocSumm \citep{Verma:17}. 
   %\textcolor{orange}{ 
   As a part of the toolkit it also provides the capability to truncate summaries to desired length. Because it relies on state-of-the-art NLP techniques for basic tasks such as tokenization and possessives, it is more accurate.
   %}
   In addition to this it provides four additional important avenues of exploration:
\begin{itemize}
% \item R normalization:  Normalizing sentence scores is done to remove bias from over-valuing a sentence based on length. NewsSumm parameterizes score normalization by exponentiating the normalizing divisor (Section~\ref{sec-norm}).
\item Normalizing Factors: Normalizing is imposed to remove bias from over-valuing a score based purely on length. For example, if a sentence is valued by the sum of its terms, then longer sentences will often receive higher scores. NewsSumm parameterizes three separate score influencing values: sentence length, and both factors of the $tfidf$ (Section~\ref{sec-norm}).
\item Integer Linear Programming (ILP): 
The problem of set-cover is a known NP-hard problem. To battle this, NewsSumm incorporates ILP to find optimal set covers (Section~\ref{sec-ilp}).
\item ILP extended: novel ILP formulations that incorporates both a word budget and alternate scoring metrics (Section~\ref{sec-budget}).
\item Title-driven summarization (Section~\ref{sec-title}: When creating titles, humans try to capture the essence of a document in a single sentence. NewsSumm leverages the title words to drive sentence selection.  
\end{itemize}
\null

\subsection{Normalizers}
\label{sec-norm}
%\textcolor{cyan}{
A common practice for selecting candidates for a final summary is to rank a sentence by the sum of its terms. Longer sentences would, on average, have higher scores. To offset this bias, ``normalization'' is used. It modifies a sentence's score based on the length of the sentence (in words). More specifically, the score of a sentence $s$, $score(s)$, is normalized by dividing by the length of $s$, $|s|$, the number of words in $s$. 
%}
% Normalization is used to balance scoring metrics by the length of the sentence. Previous work in DocSumm provided three greedy heuristics to score importance of sentences for candidate summary. In addition, it offers several parameters to tune the scores. One of these parameters, ``normalization,'' involves modifying a sentence score based on the length of the sentence (in words). 
\begin{equation}
\label{Q:norm-old}
norm(s)=\frac{score(s)}{|s|}
\end{equation}
Previous research has shown this kind of normalization can over-compensate for bias towards longer sentences \citep{lin2010multi}. NewsSumm parameterizes normalization using the idea of \citet{lin2010multi}:
\begin{equation}
\label{Q:r-norm}
\frac{score(s)}{{|s|}^r} \textrm{ , with } 0 \leq r \leq 1
\end{equation}
The original normalization parameter is Equation~\ref{Q:r-norm}, where $r = 1$.

A popular scoring heuristic is $tfidf$. $tfidf$ is a word score based on its frequency. $tf$ represents the term frequency in a document. The inverse document frequency $idf$ is computed as:
\begin{equation}
idf = log\left(\frac{N}{df}\right)
\end{equation}
Where $N$ is the number of sentences in document and $df$ is the number of sentences that include the term \citep{sparck1972statistical}.

Research has shown that the presence/absence of $tf$ and $idf$ have different affects on the final summary \citep{garciaherdandez:09}. We explore this by also parameterizing the two factors of $tfidf$. Because $tfidf$ is the product of two measures, we parameterize the influence of these factors by exponentiating them before the final product.
\begin{equation}
\label{Q:alpha-beta}
tfidf = (tf)^{\alpha} (idf)^{\beta}
\end{equation}

\subsection{Integer Linear Programming}
\label{sec-ilp}
Integer linear programming has found promise in solving computationally complex problems \citep{williams2009logic}. NewsSumm provides exploration of optimal set covers with ILP, and also a variant of ILP that includes a budget on words. One can easily extend this feature to explore set cover of ``important subsets of words'' instead of all words (or stemmed words after stopword elimination). 

Given a document D with $n$ sentences, the basic ILP formulation has an objective function to minimize the number of sentences, $s_i$. We want to minimize the following function over the vector $\hat{x}=(x_1, x_2, \dots, x_{n-1}, x_{n})$:compression and with

\begin{equation}
\sum_{i=1}^{n} x_i
\end{equation}

Each $x_i \in \hat{x}$ is a binary variable that represents inclusion/exclusion of the $i^{th}$ sentence in the summary. Then we add the dot-product constraints:

\begin{equation}\label{term-constraint}
\forall j:  \hat{x}\cdot T_j > 0
\end{equation}

Here, $T_j$ is the $j$-th row of a term-sentence matrix $T$. These constraints ensure that every word of the document is included in the summary. Note that this set of document words, can be reduced by applying stemming and removing stop words as a preprocessing step. 

\subsection{Extending ILP}
\label{sec-budget}
A variant of this ILP is to constrain it based on a budget, $L$, of the number of words allowed in the final summary. This variant then becomes a maximization problem on number of words covered. Let vector $\hat{z}=(z_1, z_2, \dots, z_{m-1}, z_{m})$ represent words covered and $z_j$ be the inclusion/exclusion of the $j^{th}$ word. Then the objective function is:

\begin{equation}\label{obj:max}
\sum_{j=1}^{m} z_j
\end{equation}

We need an integrity constraint for each term covered (i.e. if $z_j$ is included then at least one sentence with $z_j$ is also included in the summary).

% \begin{equation}
% \cancel{\forall j: \hat{x}\cdot T_j - z_j \geq 0 }
% \end{equation}

\begin{equation}
\label{eq:integrity}
\forall j: (\sum^{i}{T_{ji}}) - z_j \geq 0
\end{equation}

With only Equation~\ref{eq:integrity}, one may believe that the formulation could allow for a solution where a sentence is included, but a term of that sentence is removed. However, this is a maximization problem, so if a sentence, $x_i$ is included, then any solution which ignores a specific term will not be the maximum. Therefore, all terms will naturally be selected for inclusion. 

The only limiting factor would be the budget constraint. Let $c_i$ be the associated word count (i.e. cost) for sentence $i$, then the budget constraint is:
\begin{equation}
\sum_{i=1}^{n} x_i c_i \leq  L
\end{equation}

This constraint would force the solution summary not to exceed the budget.

We can also further modify the ILP maximization function generalizing the term-sentence matrix, $T$, into a score matrix. Rather than, each entry being a binary 0/1, we can let them be a score that represents the value of a terms inclusion in the summary. Then the goal of the ILP is to maximize this score, rather than a word count. So instead of Equation~\ref{obj:max} we would maximize the terms, $t_{ji}$ of $T$. Let $A$ be a matrix of 0/1 values such that $a_{ji}$ means inclusion of $t_{ji}$. The Equation~\ref{eq:score-max} follows as the new maximization objective function.
\begin{equation}
\label{eq:score-max}
\sum_{i=0}^{n} \sum_{j=0}^{m} (A\circ T)_{ji}
\end{equation}
Also, note that each term, $t_{j,i}$ is unique. Meaning the same word can potentially have different scores based on the sentence they appear in. It follows that the integrity constraint will change to:
\begin{equation}
\label{eq:score-int}
\forall {i,j}: a_{ji} - x_{j} = 0
\end{equation}
This constraint enforces the conditions that a sentence is selected if and only if all the words are selected.

\subsection{Title Driven Summarization}
\label{sec-title}

The goal of automatic summarization is to quickly distill the important parts of a document into a short grouping of words (or sentences). The authors of documents try to capture the essence of a document with an attention-seeking title. NewsSumm tries to leverage the authors' title, which we can think of as a  pithy summary of the document, with an unsupervised title-driven summarization algorithm as shown in Algorithm~\ref{algo::title}. In essence, a tree $T$ with sentences as nodes is created using a greedy strategy. At the root is the title sentence. It will have an edge to any sentence that overlaps with the root. The next level consists of sentences that overlap with any sentence of the previous level. Here we use a greedy strategy to connect a child sentence with the first parent sentence with which it overlaps. 
This algorithm continues until:
\begin{enumerate*}[label={(\arabic*)}]
\item all the sentences are used or
\item there are no more sentences that overlap with the tree nodes.
\end{enumerate*}

The tree structure is implicit in Algorithm~\ref{algo::title} because all we need for our title-driven summarization is the information about the level of a sentence in the tree hierarchy. 
\begin{algorithm}
\small\caption{$TitleDrivenReduction(doc, title)$}
\label{algo::title}
\begin{algorithmic}[1]
\STATE Let $G$ be ordered list of lists, where G[i] represents set of sentences at $i^{th}$ level
% \STATE $U \gets$ all words of $document$
\STATE $D \gets$ set of sentences, each sentence a set of words.
\STATE $T \gets$ set of words of $title$
\STATE $T_{next} \gets \emptyset$
\STATE $i \gets 0$
\WHILE {$D \neq \emptyset$}
	\STATE $G[i] \gets \emptyset$
	\FORALL {$S \in D$}
		\IF {$T \cap S \neq \emptyset$}
        \STATE $G[i] \gets G[i] \cup S$
        \STATE $T_{next} \gets T_{next} \cup (S - T)$
        \ENDIF
	\ENDFOR
    \IF {$T_{next} = \emptyset$}
    \STATE break
    \ENDIF
    \STATE $T \gets T_{next}$
    \STATE $T_{next} \gets \emptyset$
    \STATE $i$++
\ENDWHILE
\end{algorithmic}
\end{algorithm}

We can then order the sentences based on traversing the tree in a breadth-first search manner. This will give a new ordering of the original document, as well as potentially reducing the original document size. We believe this ordering helps to prioritize sentence ordering with the title in mind. Because of the greedy strategy implemented in  NewsSumm, there is an inherent preference to sentences visited earlier. In our title-driven summarization algorithm, we will work with the reduced set of sentences from the document, the sentences from the $TitleDrivenReduction$ tree. All the scoring algorithms will work fine, but, of course, some of the scores of the words will change, e.g. $tfidf$.
%Although this will not affect the D algorithms that scores across the whole document (i.e. tfidf), at the very least it will reduce the scope down to the sentences that have some direct/indirect overlap with the document title.
    
The $TitleDrivenReduction$ also has a depth parameter. This parameter will prematurely truncate the tree at a certain depth, i.e. $depth=1$ includes title and direct children, $depth=2$ would also include grandchildren. This version can produce a summary without using any scoring metric. It would simply create a document with sentences up to the specified depth and then truncate to summary size.

One kind of title-driven summary is to just use the $TitleDrivenReduction$ tree sentences. Another option is to run the NewsSumm algorithms on the reduced set of sentences. We explore both options below.

%%%%%%%%%%%%%%%%%%%%%%%
%%%%% EXPERIMENTS %%%%%
%%%%%%%%%%%%%%%%%%%%%%%
\section{Experiments}\label{sec:expt}
The experiments were designed to explore novel approaches to single-document extractive summarization. We begin with the datasets (Section~\ref{ssec:dataset}), pre-processing (Section~\ref{ssec:preproc}) and the ROUGE evaluation metric (Section~\ref{ssec:eval}). Next we look at the experiment design for ``Parameterization,'' ``Integer Linear Programming,'' and ``Title Driven.'' We close the section with a brief description of baselines used for comparison.
\subsection{Datasets}
\label{ssec:dataset}

\begin{table}
\small\centering
\begin{tabular}{p{.2\columnwidth} >{\centering}p{.2\columnwidth} >{\centering}p{.2\columnwidth} >{\centering}p{.2\columnwidth}}
{\bf Dataset Name} & {\bf Number of Documents} & {\bf Summaries Per Document} & {\bf Words Per Document}\tabularnewline
\hline
{\bf DUC01} & 305 & 1 & 98.35 \tabularnewline
{\bf DUC02} & 533 & 1 to 5 & 568.21 \tabularnewline
{\bf TITLE02} & 526 & 1 to 5 & 562.89 \tabularnewline
\end{tabular}
\caption{Description of datasets used in all experiments.}
\label{tab:datasets}
\end{table}

For demonstration of NewsSumm we will use  datasets from Document Understanding Conferences in 2001 and 2002.  These conferences began in 2001. After 2002, NIST discontinued the  single document extractive summarization task, because no automatic system could beat a baseline consisting of the first 100 words for news articles in a statistically significant way.
% \textcolor{blue}{Daniel - there should be a nice table here summarizing the datasets, the number of docs, the number of summaries per doc (range), words/doc. And you need to removal of duplicate docs for DUC02. }

%\textcolor{cyan}{
%An important thing to note is 
Table~\ref{tab:datasets} gives a brief overview of the datasets used. It is important to note that {\bf DUC02} {\em has 34 duplicate documents}, which could bias the results. Many previous works have ignored this issue and used  all 567 documents. We have taken care to remove all duplicate documents from {\bf DUC02}. 
%}
\begin{itemize}
\item{\bf DUC01}: These are the documents of the Document Understanding Conference (DUC). It is the test set used in 2001 \citep{duc2001info}. There are 305 documents, each with a single summary. The original set of documents had a duplicate document with a different name. So this document was removed. 
\item {\bf DUC02}: DUC documents used in test set of 2002, the last year of single-document summarization task \citep{duc2002info}. It is a collection of 533 {\em unique} news articles. 
\item {\bf TITLE02}: We also create a subset of DUC02. This dataset is the set of documents which include legitimate titles. DUC02 contains seven articles that do not have valid headlines. So these articles were excluded for the title-driven experiments (see Appendix~\ref{app:bad-headline}. 
\end{itemize}

\subsection{Pre-processing}
\label{ssec:preproc}

%\daniel{
To ensure comparable results across the methods, all documents are pre-processed in the same manner. The goal of pre-processing is to find the informative words for word scoring. We now present the pre-processing steps. 

{\bf Stemming.} Stemming removes morphological changes so that words like `musical' and `music' are the same. It also removes common verb endings to words (e.g. jump, jumps, jumped, and jumping).

{\bf Stopword Removal.} Words like `the' and `on' provide little informational content relative to main idea of a document. So a precompiled list of stopwords is removed from the document. Our software uses NLTK's English stopword list.

{\bf Punctuation.} Tokenization separates punctuation marks from the word such that they are standalone tokens. These punctuation tokens are removed from the document.

{\bf Lowercase Letters.} In all word comparisons, we always convert uppercase letters to their lowercase counterparts.
%}

\subsection{Evaluation}
\label{ssec:eval}
We use the ROUGE metric to evaluate all summaries. Given gold-standard summaries, it  is an automated metric that has been shown to have high correlation with more expensive human-based metrics \citep{lin-hovy:03}. Recent research also reveals that of the many different configurations, ROUGE-2 (bigram) correlates the best \citep{OwczarzakCDN12,Graham15} for differentiating automatic summarizers. 
In \citet{OwczarzakCDN12}, researchers show that ROUGE-2 recall has high correlation with Pyramid scores for differentiating automatic summarizers. They also show that ROUGE-1 has higher correlation with Pyramid scores for differentiating automatic versus a human summary. 

However, \citet{Graham15} recommends reporting scores with ROUGE-2 precision because it had the highest correlation with human assessments of automatic systems and it was not significantly out-performed by other evaluation techniques.

For many of our experiments, we report the \mbox{ROUGE-1} and also ROUGE-2 \mbox{F-1} scores. Unless otherwise specified all scores are F-1 scores. Some of the ROUGE-2 figures can be found in the Appendix~\ref{appendix}. For reference we have also included ROUGE-3, ROUGE-4 and ROUGE-L (longest common subsequence) in select tables. All  ROUGE evaluations were run with the parameters listed in Table~\ref{tab:rouge-params}. All summaries are truncated to 100 words before ROUGE evaluations. 

\begin{table}
\centering\small
\begin{tabular}{p{.2\columnwidth} p{.2\columnwidth} p{.4\columnwidth}}
{\bf Param} & {\bf Setting} & {\bf Detail} \\
\hline
-c & 95 & Confidence interval\\
-r & 1000 & Number samples for bootstrapping \\
-l & 100 & Upper limit on number of words in summaries \\
-m & Set & Use stemming \\
-n & 4 & ROUGE n-gram for \mbox{$1$-gram} to \mbox{$4$-gram}\\
-x & Unset & If set, ROUGE-L is not computed
\end{tabular}
\caption{Parameters used for ROUGE evaluation in all experiments.}
\label{tab:rouge-params}
\end{table}

\subsection{Optimal Parameters}\label{sec-opt}
NewsSumm's normalizing factors are compared across four preprocessing parameters. \citet{Verma:17} reports $tfidf$ option as the best performing greedy scoring metric, so we report results on $tfidf$. 
%\textcolor{orange} {
A variant of $tfidf$ is $stfidf$, which counts the term frequency per sentence rather than the whole document. 
%}
The four relevant binary parameters used are stemming (S), stopword removal (W), distinctness (D) and update on-the-fly (U). 
%For each combination of these parameters, NewsSumm generated summaries with increasing $r$-values. In addition to these, we also incorporate $r$-normalization with different ILP formulations and also in conjunction with the title-driven feature of NewsSumm.

We first look at finding the optimal $r$-values over all combinations of preprocessing options. We do a similar search for optimal $\alpha$ and $\beta$. As a final experiment a cube search over different $r$, $\alpha$ and $\beta$ values.

\subsection{Integer Linear Programming}
Integer Linear Programming lends itself to many different hard problems. NewsSumm explores different formulations to approach the automatic summarization problem (Section~\ref{sec-ilp} and Section~\ref{sec-budget}). 
% First, we look at the quality of summaries, when summarization is modeled as a pure set cover problem.  We will compare ROUGE scores of these summaries versus the whole document. We can also manually truncate these summaries to 100 words and analyze those ROUGE scores as well.
%\textcolor{red}{
We experiment with the following variations on ILP:
\begin{enumerate}
\item Budget Constraint: Forcing output to be no longer than a set threshold on number of words (e.g., 100).
\item Slack Parameter: Allowing ILP to solve with a larger budget and then truncating the result to meet the budget constraint.
\item ILP with score matrix ($score\_ilp$): Giving words a score based on $tfidf$ values. We did experiment with a variation of $tfidf$ called $stfidf$ \citep{Verma:17}, but this gave us inferior results, so we rarely show these results. 
\item Normalization: Varying $r$, $\alpha$ and $\beta$.
\end{enumerate}
%}

% NewsSumm offers a formulation of the ILP with the budget included as a restraint. We look at the quality of summaries using the budget value. NewsSumm also allows for a slack variable. This variable allows the constraint to be larger than the actual threshold, and then a simple truncation is done afterwards. We also look at how using different term scores affect the quality of the summaries. These experiments also involves exploring different combinations of $r$-value, $\alpha$ and $\beta$.

\subsection{Title Driven as Documents and Summaries}
Since title driven summaries require a title, we had to go back to the original XML version of the articles. 
This process came with its own set of hurdles:
\begin{enumerate}
\item One XML ({\em WSJ870220-0106}) had an ill-formed body (used `{\tt \&}' instead of `{\tt \&amp;}') and needed to be manually corrected.
\item Three documents (see Appendix~\ref{app:bad-headline}) did not contain any headings to start with.
\item Another four documents (see Appendix~\ref{app:bad-headline}) seemed to contain internal communications between editors and journalists in the headings. As such they did not contain relevant headings.
\item A final issue was that four document titles had no overlap with the body (see Appendix~\ref{no-overlap}). This means that the words of the title are not mentioned anywhere in the body text. For these, we implemented a failsafe where the first sentence is considered the title. This sentence along with all overlapping sentences would then be the reduced document.

\end{enumerate}
First, $TitleDrivenReduction$ was applied to the dataset to reduce documents to sentences that have direct/indirect overlap with the title. These documents were compared with the whole documents as summaries to get a sense of how much information, if any, was lost.

%\textcolor{cyan}{
We also explore different ways to produce summaries using $TitleDrivenReduction$:
\begin{itemize}
\item {\bf TITLE+DEPTH}: $TitleDrivenReduction$ applied with $depth=1$, then truncated to 100 words if needed.
\item {\bf TITLE+BFS}: $TitleDrivenReduction$ tree is traversed in breadth-first manner, and stopped after 100 words.
\item {\bf TITLE+FILTER}: $TitleDrivenReduction$ as a filter in a pipeline that also uses $score\_ilp$.
\end{itemize}
%}
% We also look at these reduced set of documents as a filter for NewsSumm algorithms. We first run NewsSumm algorithms on the original {\bf TITLE02} set and then compare the same algorithms on the reduced set. A final experiment was to use the $TitleDrivenReduction$ with $depth=1$ to create summaries. Effectively, the summary would consist of sentences that have words directly overlapping with the title. Using $depth=1$ does not guarantee a summary under the budget, so all summary sentences are put in original document order and then truncated to correct length.

\subsection{Baselines}
We use several baselines, both supervised and unsupervised, to compare our work. 
\subsubsection{Supervised Submodular}
Optimization of submodular set function has been previously used as a technique for automated summarization by \citet{sipos2013generating}, \citet{lin2010multi}, \citet{zhang2016pkusumsum} and \citet{li2012multi}. While generating a single document summary, the submodular scoring function is used to select sentences from the document taking into account maximum \textit{term coverage} while \textit{penalizing redundant sentences} in the generated summary~\citep{sipos2013generating, lin2010multi}. \citet{sipos2013generating} use trained Support Vector Machines (supervised learning system), instead of the unsupervised optimization technique proposed in ~\citet{lin2010multi}, to learn the sentence similarity as well as the word weights. 
Since the proposed summarization system was available only for multi-document summarization,\footnote{\url{http://www.cs.cornell.edu/~rs/sfour/}} we modify the system to be used for single document summarization. 
%while selecting sentences for generating an extractive summary of the source document. 
% Submodular optimization functions are used to predict an inter-sentence similarity measure based on rewarding sentences that carry the most information and at the same time penalizing sentences that are similar to each other or redundant sentences. The submodular 

Therefore, we used the submodular function proposed by ~\citet{sipos2012large}. This measure uses supervised learning to learn the best possible similarity value as well as coverage/redundancy trade-off from the training data that is fed into it. This supervised system proposed learns a `\textit{parameterized monotone submodular scoring}' function from the dataset of news articles and their summaries when given to it.

%\textcolor{orange}{
We also integrate the supervised submodular approach with our title-driven algorithms. For direct comparison, we report all results only on the {\bf TITLE02} data and with five-fold cross-validation. We bin the 526 documents of {\bf TITLE02} into bins of size 105, 105, 105, 105 and 106. Following is a list of experiments done with submodular approach:
\begin{itemize}
\item Using supervised submodular approach by itself. This provides a simple baseline for comparison.
\item First using $TitleDrivenReduction$ on each document to reduce number of sentences for each document. Then applying the supervised submodular approach.
\item Using $TitleDrivenReduction$ with $depth=1$, followed by the supervised submodular approach.
% \item Let supervised submodular approach produce full summaries and then apply $TitleDrivenReduction$ on these summaries with truncation.
\end{itemize}
%}

\subsubsection{PKUSUMSUM}
For comparison, we have run PKUSUMSUM on {\bf DUC02}. The tool implements several previous works on summarization. The five methods we have run are as follows:
\begin{itemize}
\item {\bf Lead} is a baseline method that uses initial sentences of the document as summary.
\item {\bf Centroid} \citet{RadevJST04} scores sentences using a centroid. The centroid is a pseudo-sentence to represent a group of documents. Sentences are scored based on similarity to this centroid, sentence position and similarity to first sentence of a document.
\item {\bf TextRank} \citet{mihalcea:textrank04} is a graph based method that has sentences as vertexes and edge weights based on overlap. Random walk is used to determine sentence ranks.
\item {\bf LexPageRank} \citet{erkan2004lexrank} is also a graph based method. Like TextRank its vertexes are the sentences. However, the edges are based on cosine similarity. Page rank is run on this graph and the node values are the ``LexPageRank'' values used for a greedy choice.
\item {\bf Unsupervised Submodular} Uses algorithms in \citet{lin2010multi,li2012multi}. These are {\em unsupervised methods} based on submodular set functions in mathematics.
\end{itemize}

%%%%%%%%%%%%%%%%%%%%%%%%%%%%%%%%%%
%%%%% RESULTS AND DISCUSSION %%%%%
%%%%%%%%%%%%%%%%%%%%%%%%%%%%%%%%%%
\section{Results and Discussion}
\label{sec:results}
We now present the results of our experiments and discuss their significance.
\begin{table*}
	\centering\small
	\setlength{\tabcolsep}{3pt}
	\begin{tabular}{cccc | c c c c | c c c c | c c c c}
	\hline
    \multicolumn{4}{c}{} & \multicolumn{4}{c}{{\bf DUC01}} & \multicolumn{4}{c}{{\bf DUC02}} & \multicolumn{4}{c}{{\bf TITLE02}}\\
    \hline
    \multicolumn{4}{c|}{Params Used} & \multicolumn{2}{c}{$r$-Value Ends} & Best & ROUGE-1 & \multicolumn{2}{c}{$r$-Value Ends} & Best & R-1 & \multicolumn{2}{c}{$r$-Value Ends} & Best & R-1 \\
    S & W & D & U & 0.0 & 1.0 & $r$ & F-1 score & 0.0 & 1.0 & $r$ & F-1 score & 0.0 & 1.0 & $r$ & F-1 score \\
    \hline
  &   &   &   & 0.4030 & 0.4065 & 0.7 & 0.4122 & 0.4255 & 0.4242 & 0.7 & 0.4291 & 0.4260 & 0.4242 & 0.7 & 0.4291\\
X &   &   &   & 0.4022 & 0.4116 & 0.8 & 0.4201 & 0.4274 & 0.4267 & 0.7 & 0.4321 & 0.4278 & 0.4269 & 0.7 & 0.4322\\
  & X &   &   & 0.4105 & 0.3980 & 0.5 & 0.4146 & 0.4360 & 0.4162 & 0.5 & 0.4375 & 0.4370 & 0.4170 & 0.5 & 0.4382\\
  &   & X &   & 0.4110 & 0.3934 & 0.7 & 0.4160 & 0.4330 & 0.4173 & 0.6 & 0.4444 & 0.4338 & 0.4177 & 0.6 & 0.4451\\
  &   &   & X & 0.3918 & 0.3996 & 0.6 & 0.4007 & 0.4238 & 0.4213 & 0.0 & 0.4238 & 0.4239 & 0.4216 & 0.0 & 0.4239\\
X & X &   &   & 0.4111 & 0.4012 & 0.5 & 0.4186 & 0.4366 & 0.4192 & 0.4 & 0.4393 & 0.4376 & 0.4199 & 0.4 & 0.4398\\
X &   & X &   & 0.4145 & 0.3988 & 0.7 & 0.4195 & 0.4346 & 0.4204 & {\em\bf 0.6} & {\em\bf 0.4478} & 0.4350 & 0.4208 & {\em\bf 0.6} & {\em\bf 0.4481}\\
X &   &   & X & 0.3935 & 0.4010 & 0.8 & 0.4029 & 0.4219 & 0.4228 & 0.2 & 0.4239 & 0.4219 & 0.4232 & 0.2 & 0.4240\\
  & X & X &   & 0.4144 & 0.4022 & 0.5 & 0.4219 & 0.4407 & 0.4228 & 0.4 & 0.4446 & 0.4412 & 0.4234 & 0.5 & 0.4453\\
  & X &   & X & 0.3991 & 0.3926 & 0.5 & 0.4017 & 0.4228 & 0.4192 & 0.2 & 0.4238 & 0.4232 & 0.4198 & 0.2 & 0.4244\\
  &   & X & X & 0.3962 & 0.3966 & 0.8 & 0.4029 & 0.4261 & 0.4209 & 0.5 & 0.4310 & 0.4258 & 0.4212 & 0.5 & 0.4313\\
X & X & X &   & 0.4161 & 0.4054 & {\em\bf 0.6} & {\em\bf 0.4248} & 0.4428 & 0.4266 & 0.5 & 0.4474 & 0.4436 & 0.4273 & 0.5 & 0.4479\\
X & X &   & X & 0.3974 & 0.3958 & 0.5 & 0.4006 & 0.4250 & 0.4223 & 0.5 & 0.4261 & 0.4257 & 0.4229 & 0.5 & 0.4268\\
X &   & X & X & 0.3986 & 0.3965 & 0.6 & 0.4041 & 0.4259 & 0.4214 & 0.6 & 0.4293 & 0.4255 & 0.4217 & 0.6 & 0.4296\\
  & X & X & X & 0.4002 & 0.3986 & 0.4 & 0.4082 & 0.4295 & 0.4289 & 0.4 & 0.4325 & 0.4301 & 0.4294 & 0.4 & 0.4329\\
X & X & X & X & 0.3985 & 0.3980 & 0.5 & 0.4084 & 0.4318 & 0.4277 & 0.5 & 0.4374 & 0.4325 & 0.4284 & 0.5 & 0.4383\\
	\end{tabular}
    \caption{Best $r$-values for each set of parameters (explained in Section~\ref{sec-opt})  using $tfidf$ Greedy Heuristic on {\bf DUC02}. Best overall ROUGE-1 F-1 scores for each dataset are marked in {\em\bf bold italics}. }
    \label{tab:rnorm}
\end{table*}

\subsection{Importance of R-Normalization}
\label{ssec:rnorm}
Table~\ref{tab:rnorm} reports the best $r$-value found for all combinations of binary parameters for the $tfidf$ metric. For all three datasets, the optimal value is never $r = 0$ or $r = 1$. Even with a simple metric like $tfidf$, we see the value of parameterizing the sentence normalization. These findings are in line with \citep{lin2010multi}. The value of normalization based on size is not always beneficial, but by adjusting the influence of the normalization (i.e. $r$-value), the results show that the benefits of normalization can always be leveraged.

\subsection{ILP with Budget Constraint}
\label{sec-res-budget}
This algorithm also incorporates a budget in the search for best coverage. In those experiments the algorithm preprocessed the input text with stemming and stopword removal. The scores show that ILP fails to overcome the ROUGE scores of $tfidf$ based greedy heuristic on {\bf DUC02}.

A possibility is that a 100 word {\em initial} threshold is over-constraining the search for good set covers. So further runs were done with increasing amounts of budget (called threshold in the table) and then a truncation to 100 words before measuring the ROUGE scores. All these results are reported in Table~\ref{summ:truncate}. 
Interestingly, adding more and more slack to Budget-ILP keeps improving the scores, until it matches the ROUGE-1 F-1 score of the Lead baseline of PKUSUMSUM (compare with Table~\ref{tab:pku-duc02}). We expect it to match the ROUGE-2 and ROUGE-L scores as well if the slack is increased further.  

\begin{table}
\small
\begin{tabular}{c c c c}
\hline
& \multicolumn{3}{c}{\bf ROUGE METRIC}\\
\cline{2-4}
{\bf Threshold} & ROUGE-1 & ROUGE-2 & ROUGE-L \\
\hline
100 & 0.4083 & 0.1512 & 0.3430 \\
110 & 0.4125 & 0.1557 & 0.3475 \\
120 & 0.4129 & 0.1586 & 0.3492 \\
130 & 0.4143 & 0.1592 & 0.3500 \\
140 & 0.4172 & 0.1629 & 0.3531 \\
150 & 0.4197 & 0.1646 & 0.3558 \\
\dots & \dots & \dots & \dots \\
950 & 0.4764 & 0.2244 & 0.4123 \\
\end{tabular}
\caption{Threshold effect on $budget\_ilp$ on {\bf DUC02} with truncation.}
\label{summ:truncate}
\end{table}

\subsection{ILP with Score}\label{res:score-ilp}
\begin{figure}
	\centering
    \includegraphics[width=\columnwidth]{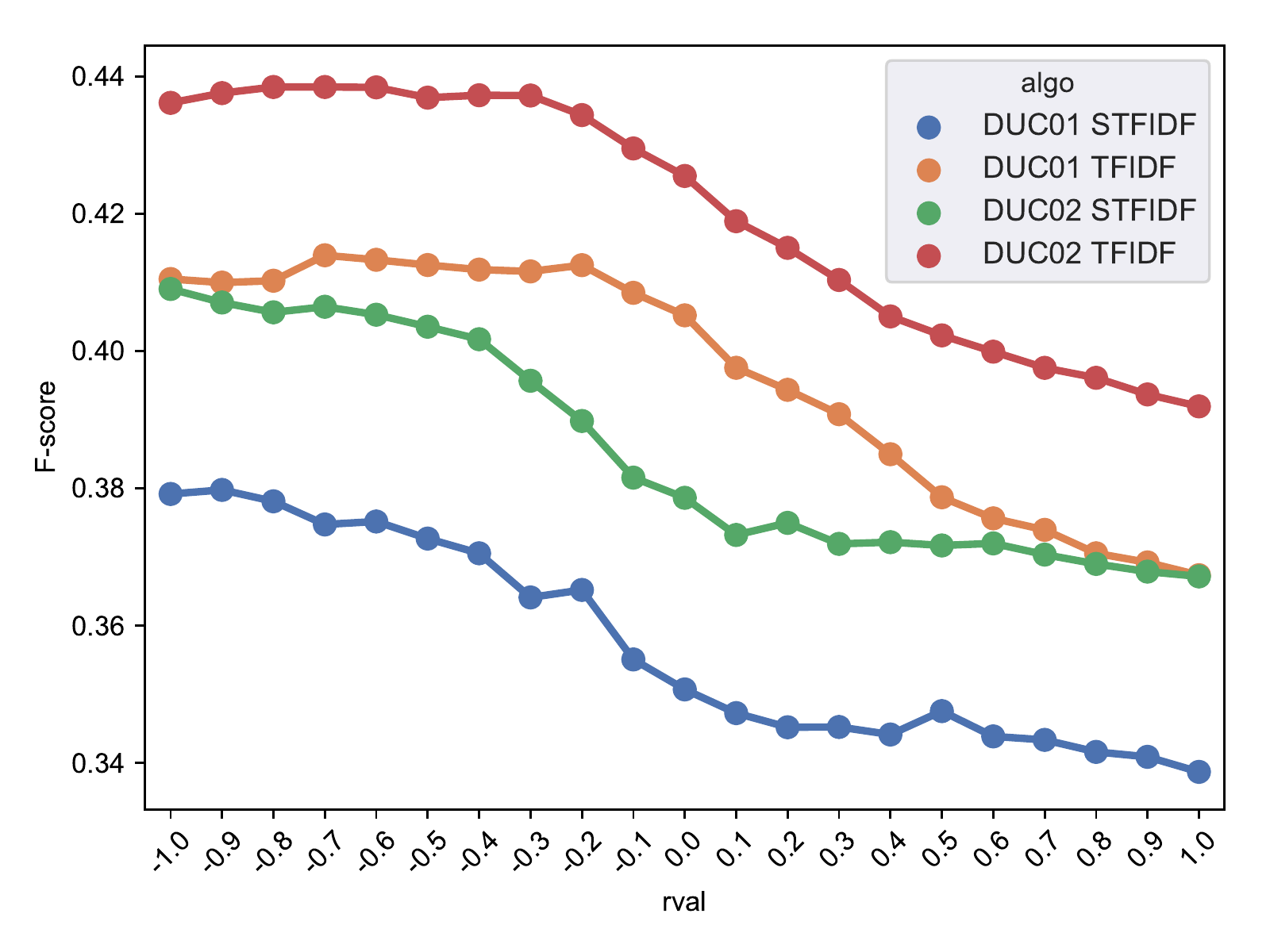}
    \caption{Line plots of ROUGE-1 F-1 scores across different $r$-values with $score\_ilp$.}
    \label{fig:tf2stf-rouge}
\end{figure}

%\textcolor{red}{
    ILP with a {\em strict} budget constraint fails to do better then greedy $tfidf$ heuristic. We believe this is because the formulation is giving equal weight to all words. To investigate this further, we apply a score to each term. Rather than giving equal value to all words, we introduce a scoring matrix. 
%}
When the ILP is run using $tfidf$ values for the scoring metric it comes closer to the best pure greedy $tfidf$ run, with a ROUGE-1 score of $0.4266$ on {\bf DUC02}. Using the $tfidf$ score for each term proves to be the best scoring metric for this version of ILP. This is competitive with the state-of-the-art in unsupervised extractive summarization. We also add r-normalization and found that this detracted from the scores. These are seen in the values of $r>0$  in Figure~\ref{fig:tf2stf-rouge}. 
    Initially, the experiments only looked at values in the range $[0.0, 1.0]$. Figure~\ref{fig:tf2stf-rouge} shows there is a trend of increasing scores as $r$ tends to $0.0$. Hence, we generated summaries using negative $r$-values and found that it does improve the ROUGE-1 scores.

\begin{table}
\small
\begin{tabular}{c c c c}
\hline
& \multicolumn{3}{c}{\bf ROUGE METRIC}\\
\cline{2-4}
{\bf Threshold} & ROUGE-1 & ROUGE-2 & ROUGE-L \\
\hline
100 & 0.4391 & 0.1870 & 0.3782 \\
110 & 0.4414 & 0.1888 & 0.3793 \\
120 & 0.4463 & 0.1936 & 0.3836 \\
130 & 0.4540 & 0.2007 & 0.3912\\
140 & 0.4541 & 0.2021 & 0.3919 \\
150 & 0.4548 & 0.2002 & 0.3920 \\
% x100 & 0.4413 & 0.1890 & 0.3798 \\
% x110 & 0.4467 & 0.1936 & 0.3836 \\
% x120 & 0.4478 & 0.1950 & 0.3850 \\
% x130 & 0.4526 & 0.1998 & 0.3894\\
% x140 & 0.4541 & 0.2029 & 0.3922 \\
% x150 & 0.4547 & 0.2014 & 0.3921 \\
% x160 & 0.4596 & 0.2074 & 0.3973 \\
% x170 & 0.4606 & 0.2081 & 0.3980 \\
% x180 & 0.4611 & 0.2097 & 0.3993 \\
% x190 & 0.4626 & 0.2099 & 0.4005 \\
% x200 & 0.4643 & 0.2130 & 0.4031 \\
\dots & \dots & \dots & \dots \\
900 & 0.4780 & {\bf 0.2260} & 0.4139 \\
910 & 0.4780 & 0.2259 & 0.4138 \\
920 & {\bf 0.4782} & {\bf 0.2260} & {\bf 0.4141} \\
930 & 0.4781 & {\bf 0.2260} & 0.4139 \\
940 & 0.4779 & 0.2259 & 0.4137 \\
950 & 0.4776 & 0.2256 & 0.4133 \\
960 & 0.4779 & 0.2258 & 0.4137 \\
970 & 0.4779 & 0.2257 & 0.4136 \\
980 & 0.4778 & 0.2257 & 0.4135 \\
990 & 0.4778 & 0.2258 & 0.4135 \\
1000 & 0.4779 & 0.2286 & 0.4136 \\
1010 & 0.4779 & 0.2257 & 0.4136 \\
1020 & 0.4776 & 0.2257 & 0.4134 \\
1030 & 0.4775 & 0.2257 & 0.4133 \\
1040 & 0.4775 & 0.2256 & 0.4133 \\
1050 & 0.4775 & 0.2255 & 0.4133 \\
% x900 & 0.4775 & 0.2256 & 0.4132 \\
% x910 & 0.4778 & 0.2258 & 0.4137 \\
% x920 & 0.4778 & 0.2258 & 0.4136 \\
% x930 & 0.4775 & 0.2256 & 0.4133 \\
% x940 & 0.4776 & 0.2257 & 0.4134 \\
% x950 & 0.4778 & 0.2258 & 0.4135 \\
% x960 & 0.4779 & 0.2260 & 0.4136 \\
% x970 & 0.4776 & 0.2255 & 0.4134 \\
% x980 & 0.4777 & 0.2256 & 0.4134 \\
% x990 & 0.4776 & 0.2257 & 0.4133 \\
% x1000 & 0.4775 & 0.2256 & 0.4132 \\
% x1050 & 0.4774 & 0.2255 & 0.4131 \\
\dots & \dots & \dots & \dots \\
2350 & 0.4771 & 0.2251 & 0.4129 \\

\end{tabular}
\caption{Threshold effect on $score\_ilp$ with $tfidf$ on {\bf DUC02} with truncation. Bolding shows highest scores in each column.}
\label{summ:truncate}
\end{table}

We believe an explanation can be found in the formulation of $tfidf$. The $tfidf$ includes a form of normalization, by penalizing a word that appears often through a corpus. Because r-normalization serves a similar goal, it is over penalizing words. A negative $r$-value serves to remove this over-penalizing by crediting scores that come from longer sentences. 
%\textcolor{orange}{
Figure~\ref{fig:tf2stf-rouge} shows that negative $r$-values indeed improve the score. The $stfidf$ scores in Figure~\ref{fig:tf2stf-rouge} suggest that more improvement can be found beyond an $r$-val of $-1.0$. However additional experiments in that range did not improve, and instead showed a steady decline. The best score is seen with $tfidf$ on {\bf DUC02} with an F-1 score of $0.4385$ found at $r = -0.8$.
%} 

\begin{figure}
	\centering
    \includegraphics[width=\columnwidth]{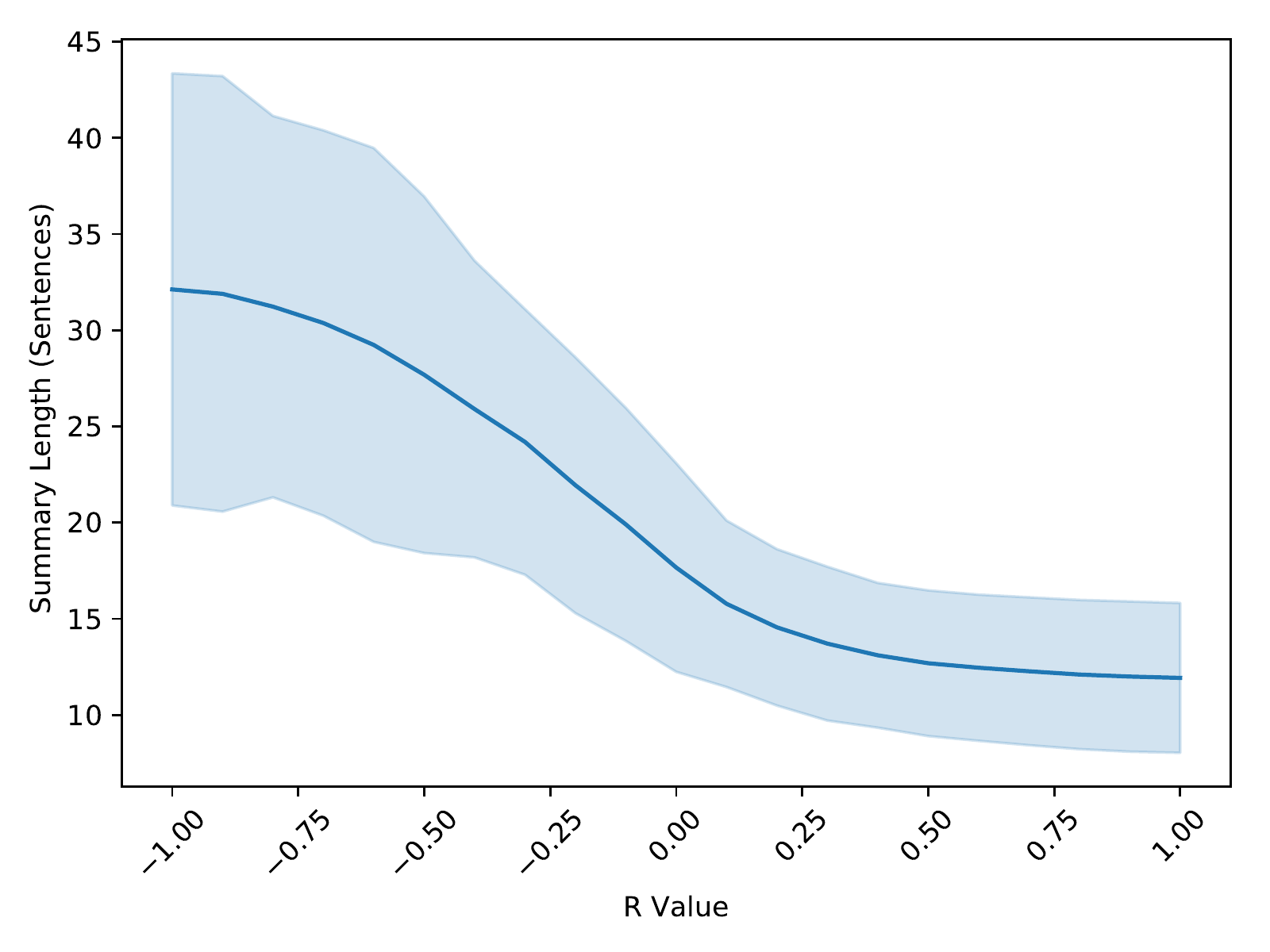}
    \caption{Average sentence lengths for summaries with standard deviation on {\bf DUC01} with $score\_ilp$ with $tfidf$ scores.}
    \label{fig:sent-len-std01}
\end{figure}

\begin{figure}
	\centering
    \includegraphics[width=\columnwidth]{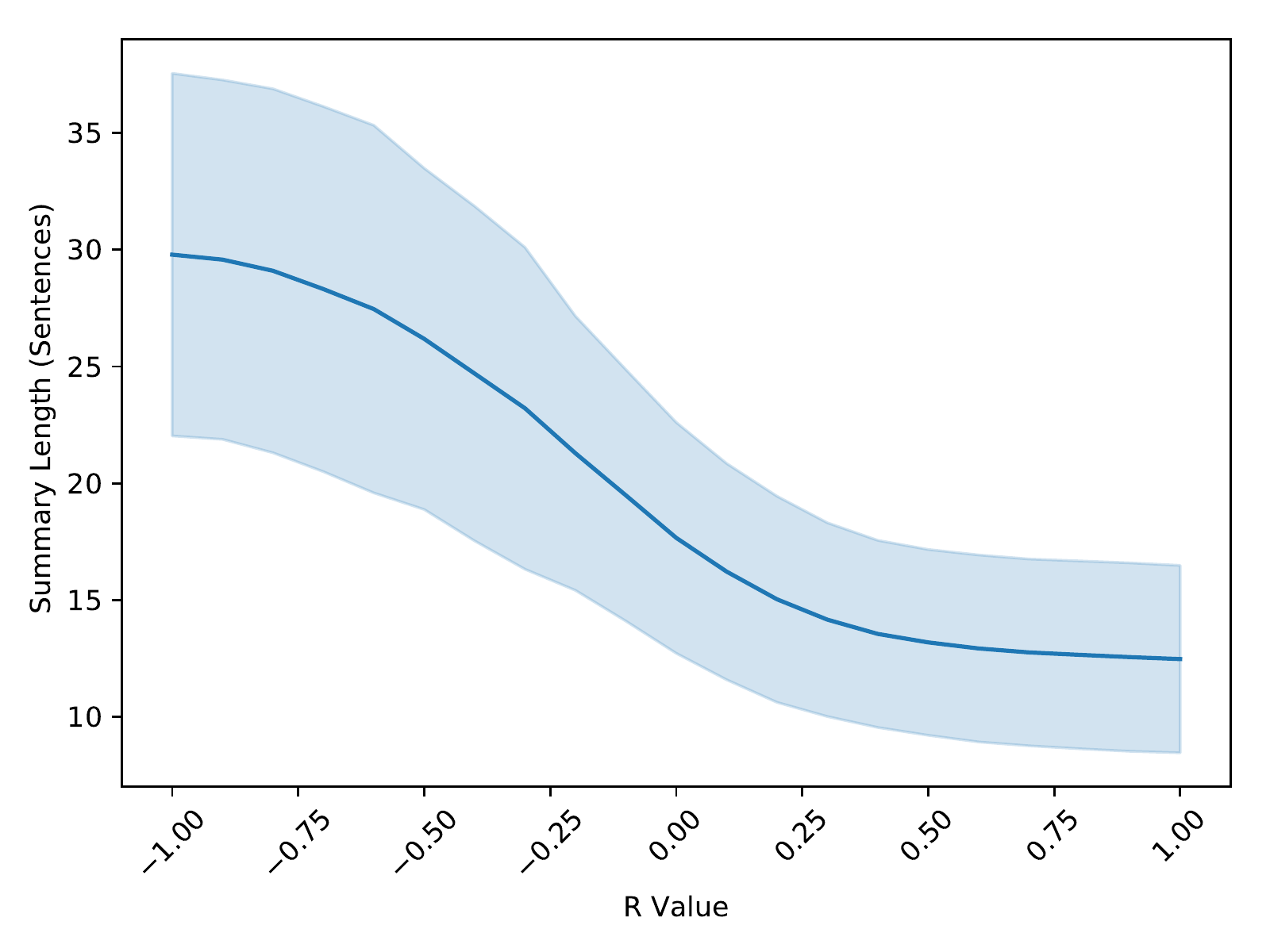}
    \caption{Average sentence lengths for summaries with standard deviation on {\bf DUC02} with $score\_ilp$ with $tfidf$ scores.}
    \label{fig:sent-len-std02}
\end{figure}

Since normalization influences the contribution of sentence length to the score of a sentence, negative $r$-values would favor longer sentences.  %\textcolor{red}{
    We plotted the average sentence length along with standard deviation for all the summaries at different $r$ values. The trend can be seen in Figure~\ref{fig:sent-len-std01} and Figure~\ref{fig:sent-len-std02} for {\bf DUC01} and {\bf DUC02}, respectively. However, the standard deviation depicted by the bands show that as the average sentence length decreases so does the standard deviation. This means that although negative $r$-values produce summaries that include longer sentences, the algorithm does not exclusively choose longer sentences. Another interesting point is the similar shape of the sentence length curve and the F-score curve. We computed the Pearson Correlation as $0.939$. Expressed differently, it means there is a strong correlation between higher ROUGE scores and longer sentences.
%}

\begin{figure}
\centering
\includegraphics[width=\columnwidth]{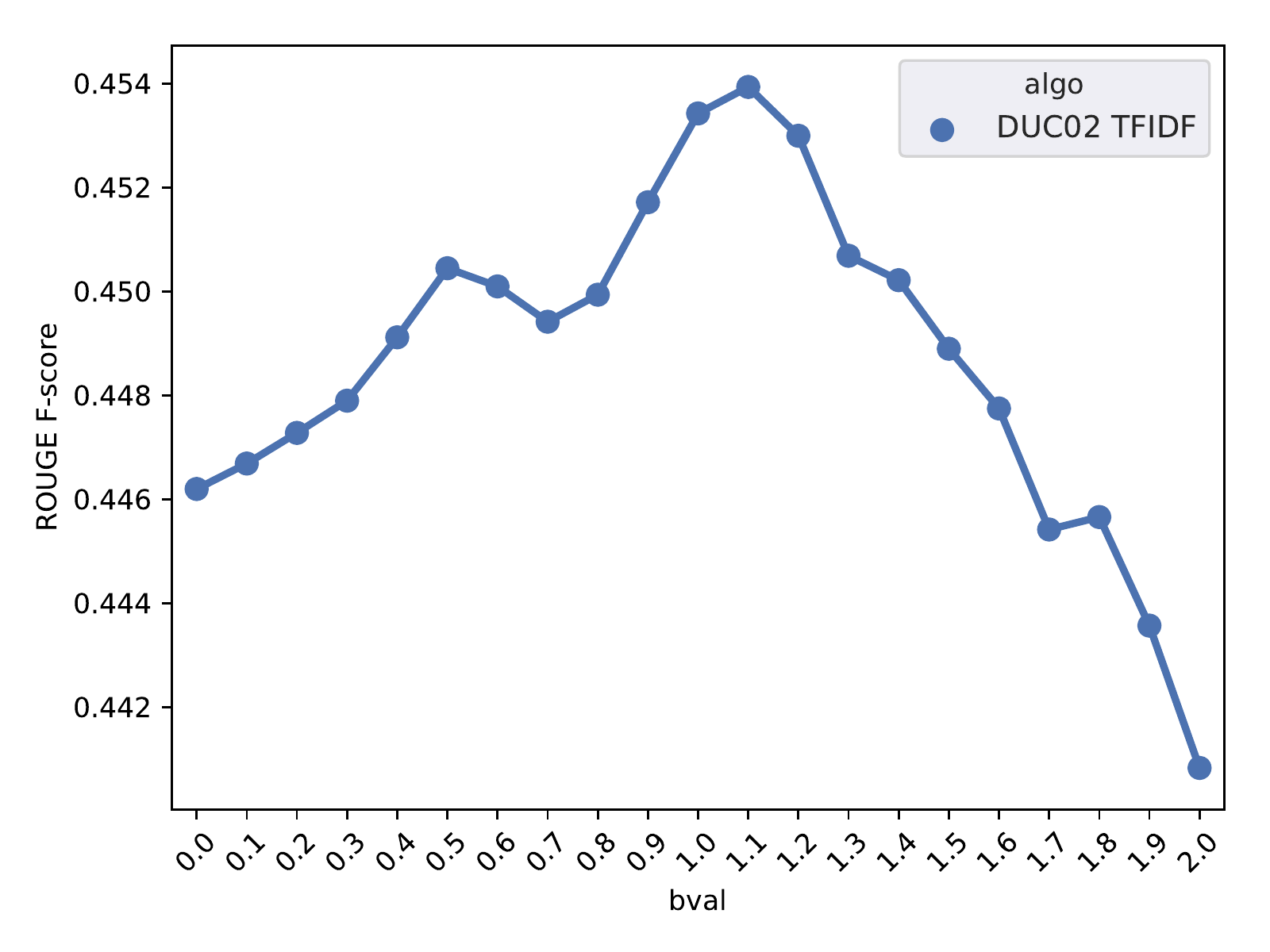}
\caption{{\bf DUC02} ROUGE-1 F-1 scores for varying $\beta$ after finding best $\alpha$ ($ = 1.1$) when $r = -0.4$ (Uses $score\_ilp$.)}
\label{fig:progressive}
\end{figure}

%\textcolor{red}{
Next we look at the influence of the two factors of $tfidf$, namely $tf$ and $idf$. For the initial experiment, we fixed $r$-value at $-0.4$ (the best ROUGE-1 from previous experiment) and ran $score\_ilp$ with varying $\alpha$ values from $-1.0$ to $1.0$ in increments of $0.1$. The best $\alpha$ was found to be at $1.0$. The scores showed an increasing trend so an additional experiment was done to search in range between $1.0$ and $2.0$. Here we found that the score continues to improve until $\alpha = 1.1$, after which there is a quick decline. We then fix $\alpha = 1.1$ and vary $\beta$ in same range of values, we see that the best $\beta$ is $\beta = 1.1$ with a ROUGE-1 F-score of $0.4539$ (Figure~\ref{fig:progressive}). Effectively, this means that $\beta$ does not induce a better solution. We investigate this in following experiment.
%}

\begin{figure}
    \centering
    \includegraphics[width=\columnwidth]{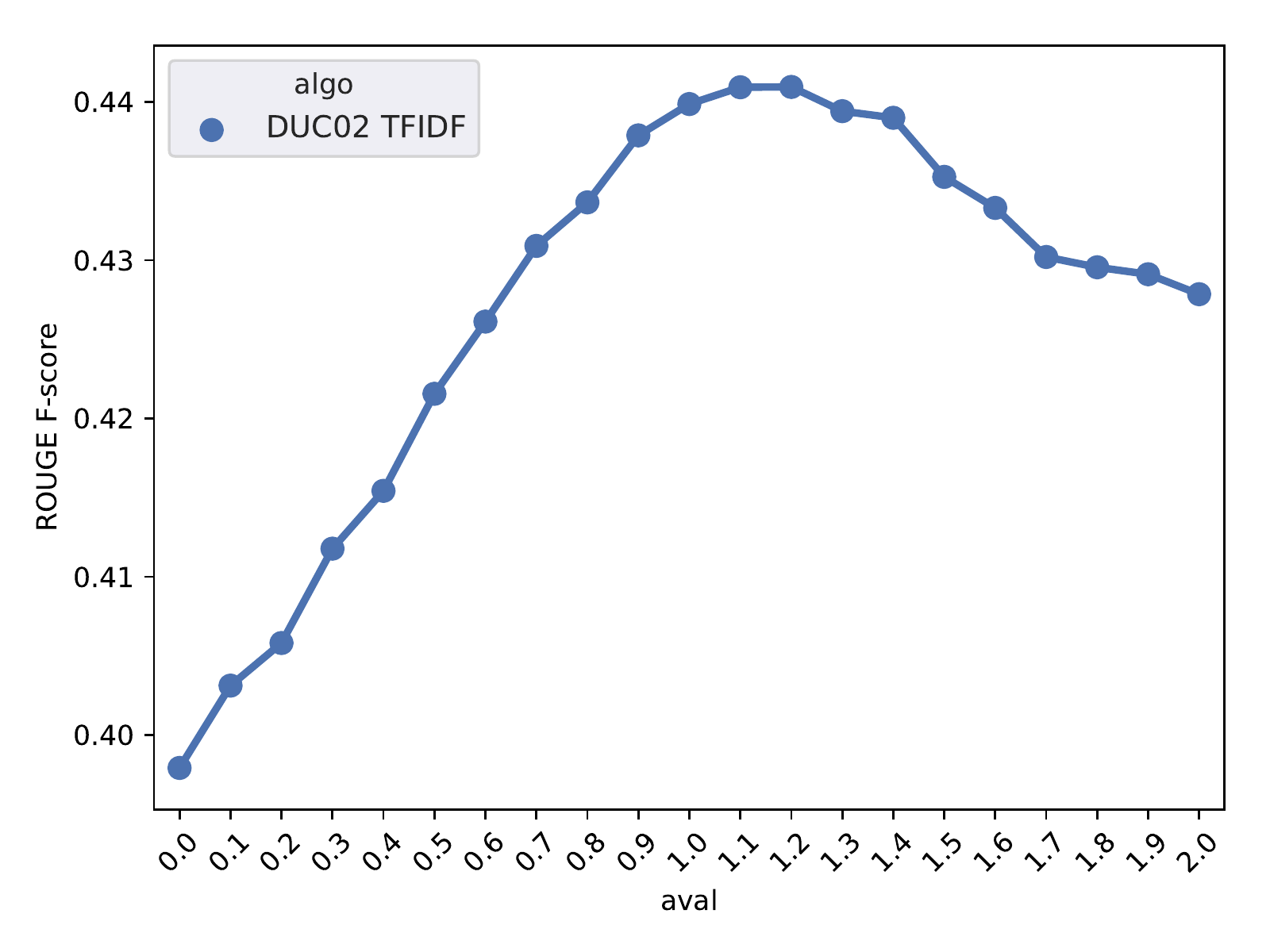}
    \caption{Effect of $\alpha$ ($\beta = 1$, $r=0$) on ROUGE F-1 score on {\bf DUC02} using $score\_ilp$.}
    \label{fig:alpha-rouge}
\end{figure}

\begin{figure}
    \centering
    \includegraphics[width=\columnwidth]{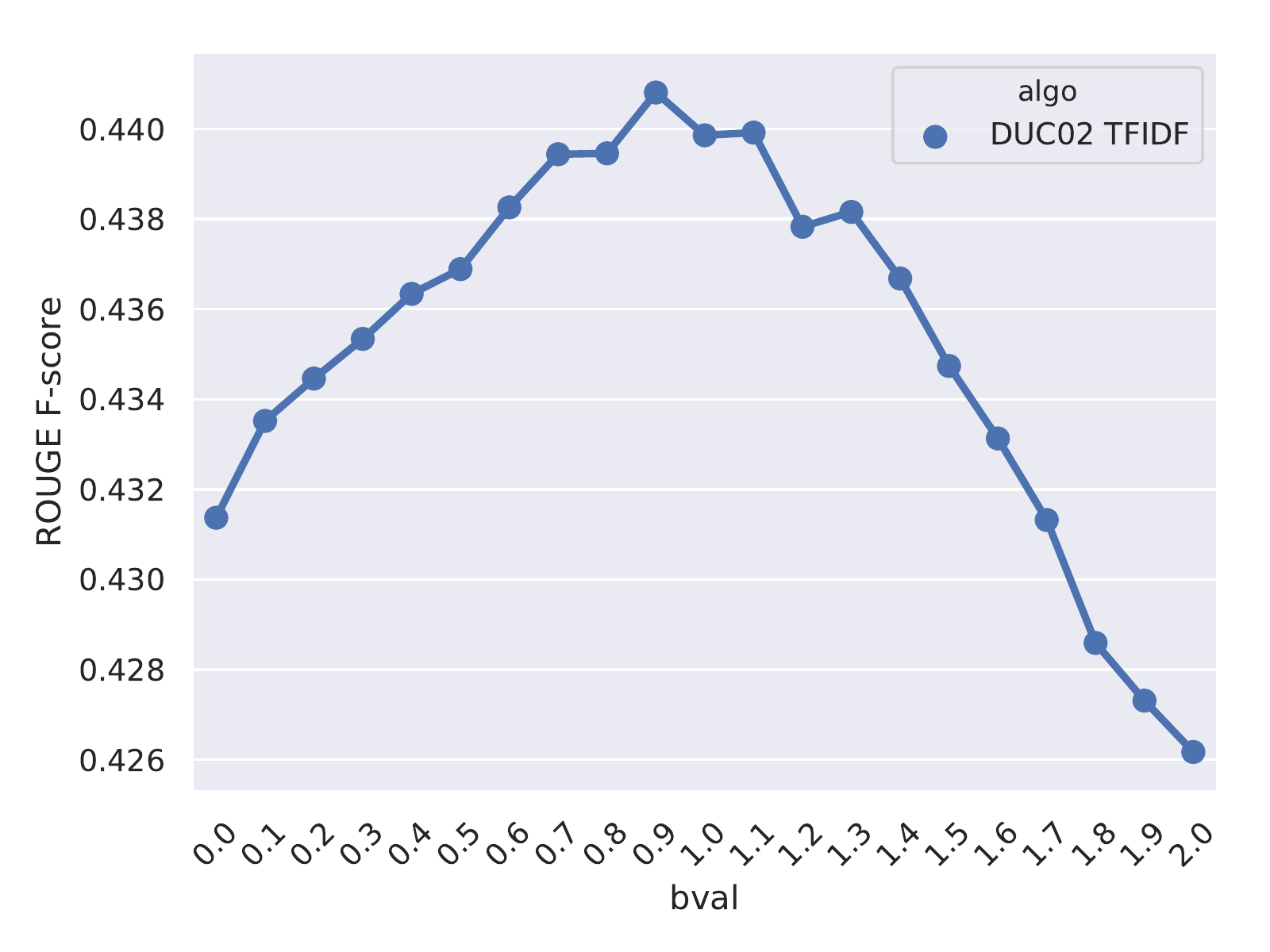}
    \caption{Effect of $\beta$ ($\alpha=1$, $r=0$) on ROUGE F-1 score on {\bf DUC02} using $score\_ilp$.}
    \label{fig:beta-rouge}
\end{figure}

\begin{figure}
\centering
\includegraphics[width=\columnwidth]{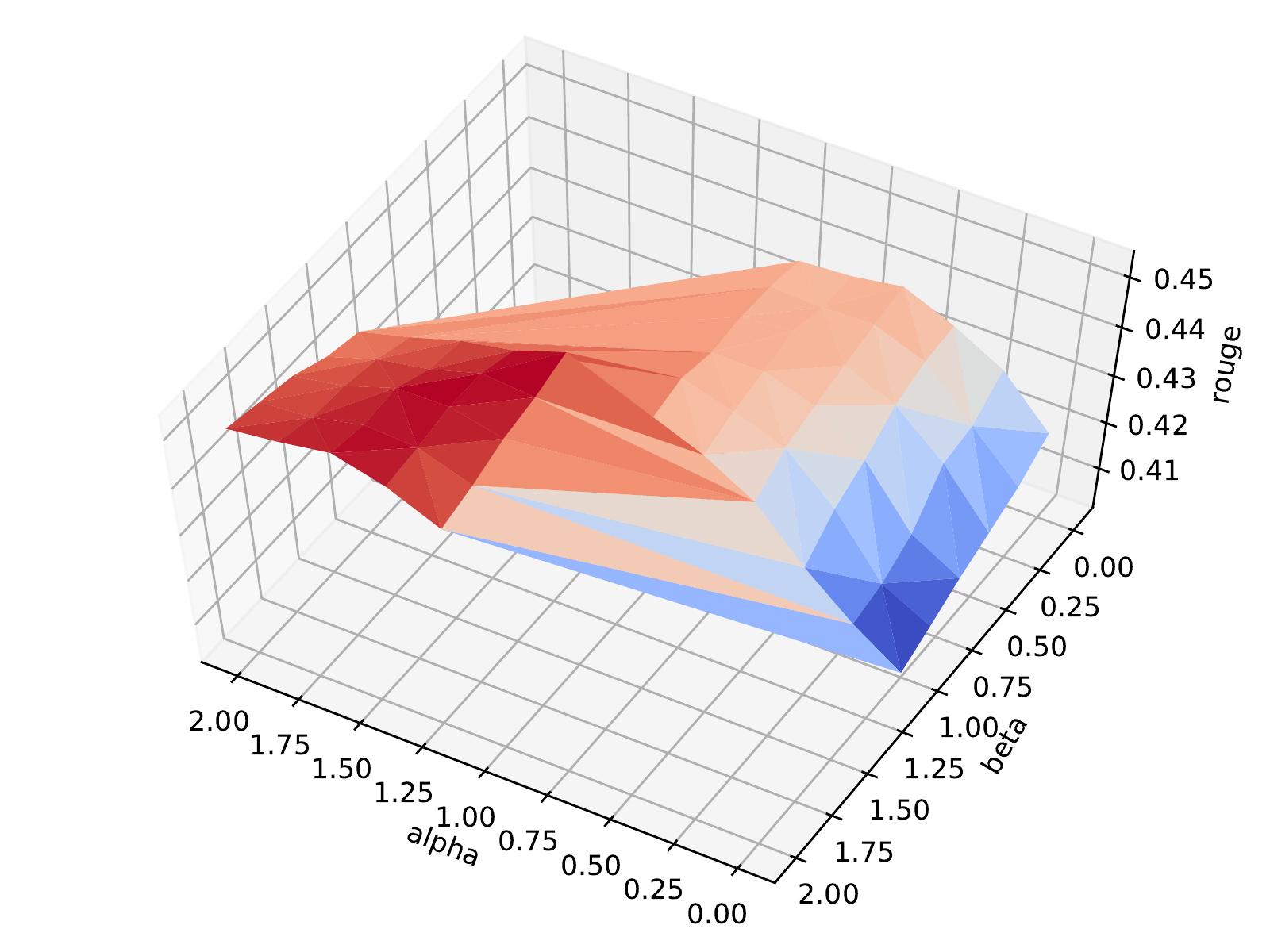}
\caption{ROUGE-1 F-1 scores with varying $\alpha$ and $\beta$ values with $r=-0.4$ (the best $r$-value from cube search) on {\bf DUC02} using $score\_ilp$. (We used a slack variable of 130, because this showed the largest boost in preliminary experiments).}
\label{fig:cubesearch}
\end{figure}

Because of the local max found at $\beta=0.7$, there is reason to believe that different combinations of $\alpha$ and $\beta$ may produce superior summaries. To further explore this we do three separate experiments:
\begin{enumerate}
    \item Only vary $\alpha$ ($\beta=1$, $r=0$) (Figure~\ref{fig:alpha-rouge}).
    \item Only vary $\beta$ ($\alpha=1$ and $r=0$) (Figure~\ref{fig:beta-rouge}).
    \item A cube search across the $r$-value, $\alpha$ and $\beta$.  We show the three
    dimensional plot for the best $r$-value (Figure~\ref{fig:cubesearch}). 
\end{enumerate}
Note that default values for $\alpha$ and $\beta$ are $1.0$, because they are part of the tfidf metric. This basically leaves them in a neutral state with regards to the tfidf metric, i.e. when the studied parameter is also $1.0$ we should see the result as if the tfidf score was untouched. 

Varying $\alpha$ and $\beta$ individually does affect the ROUGE scores. The best $\alpha$ is at $1.1$ and $\beta$ at $1.1$.
% values show an overall decreasing trend (Figure~\ref{fig:beta-rouge}).

The cube search showed that the maximum ROUGE-1 scores occurred at $r = -0.4$. So, Figure~\ref{fig:cubesearch} shows a 3D surface plot with this $r$ value. The maximum ROUGE-1 score of $0.4540$ is found at $\alpha=1.2$ and $\beta=1.2$. This is different from before, so the cube search did find higher ROUGE scores. Similar results were found when the cube search was done on {\bf TITLE02}, its results can be found in Appendix~\ref{appendix} (Figures \ref{cubesearch-duc02-r2}, \ref{cubesearch-title02} and \ref{cubesearch-title02-r2}).

Using the best results found in the cube search, we now revisit the effect of threshold on $score\_ilp$. Like, the $budget\_ilp$, Table~\ref{summ:truncate} reveals that $score\_ilp$ also gradually improves in score. However, we see that it begins to do better than the lead method. And furthermore, we also see that it peaks at a threshold amount of 960, giving a score of $0.4779$.

\subsection{Title Driven as Documents and Summaries}\label{res:title}

NewsSumm uses the $TitleDrivenReduction$ algorithm to reduce a document to sentences that are indirectly or directly overlapping with the title of the document. To evaluate the benefit of this, we first run all the non-ILP algorithms on the reduced dataset, {\bf TITLE02} (results in Table~\ref{tab:rnorm}). Nearly all the scores were boosted and the best result was a ROUGE-1 F-1 Score of $0.4483$.

\begin{table}
% ../results/2018-0328-0428/rouge
\centering
\begin{tabular}{lccc}
Metric & Recall & Precision & F-Score\\
\hline
ROUGE-1 & 0.9027 & 0.2041 & 0.3214\\
ROUGE-2 & 0.5519 & 0.1267 & 0.1990\\
ROUGE-3 & 0.3692 & 0.0864 & 0.1352\\
ROUGE-4 & 0.2689 & 0.0635 & 0.0992\\
ROUGE-L & 0.7842 & 0.1775 & 0.2794\\
% ROUGE-1 & 0.9028 & 0.2041 & 0.3218\\
% ROUGE-2 & 0.5522 & 0.1269 & 0.1994\\
% ROUGE-3 & 0.3703 & 0.0868 & 0.1359\\
% ROUGE-4 & 0.2705 & 0.0640 & 0.1000\\
% ROUGE-L & 0.7907 & 0.1789 & 0.2820\\
% ROUGE-W-1.2 & 0.3017 & 0.1149 & 0.1592\\
\end{tabular}
\caption{Using {\bf TITLE02} documents as summaries.}
\label{tab:docassumm}
\end{table}

% \begin{table}[t]
% % ../results/2018-0328-0428/rouge
% \centering
% \begin{tabular}{lccc}
% Metric & Recall & Precision & F-Score\\
% \hline
% ROUGE-1 & 0.89429 & 0.20125 & 0.31779\\
% ROUGE-2 & 0.54669 & 0.12500 & 0.19677\\
% ROUGE-3 & 0.36637 & 0.08539 & 0.13392\\
% ROUGE-4 & 0.26761 & 0.06295 & 0.09854\\
% ROUGE-L & 0.78316 & 0.17645 & 0.27852\\
% ROUGE-W-1.2 & 0.30173 & 0.11487 & 0.15922\\
% \end{tabular}optimal
% \caption{Using {\bf TITLE02} with documents as summaries.}
% \label{tab:docassumm}
% \end{table}

{\bf Limit on ROUGE Recall.} In addition to this experiment we also look at how these reduced documents do when viewed as a summary themselves. ROUGE evaluates by finding the word overlap between the model summary and generated summary. Because the model summary were created by human annotators, they are abstractive and will often make use of words not in the actual article. This puts an inherent limit to the highest achievable score, when the summary can only include words from the document (extractive summaries). To find this limit we run the ROUGE {\em on the documents themselves as summaries} evaluation against the human summaries. Of course the precision will be very low, but the recall will show the maximum coverage possible of the human summary as evaluated by ROUGE. That result is presented in Table~\ref{tab:docassumm}. In \citet{Verma:17}, researchers report the ROUGE-1 Recall as $0.907$ for full documents as summaries, and we see that title driven summaries do not detract much from that limit.

{\bf TITLE+DEPTH}. $TitleDrivenReduction$ by itself is used to create a summary. This is done by simply looking only at the sentences that are directly overlapping with the title (i.e. $depth=1$). If these sentences go over the 100 word budget, then the summary is truncated to fit. The results are in Table~\ref{table:dep1}. We see that this produces a competitive ROUGE score with $score\_ilp$ with $tfidf$ with an F-1 score of $0.4745$. This further confirms that the title holds salient information with respect to a summary. 

\begin{table}
\centering
\begin{tabular}{lccc}
Metric & Recall & Precision & F-Score\\
\hline
ROUGE-1 & 0.4719 & 0.4778 & 0.4745\\
ROUGE-2 & 0.2201 & 0.2228 & 0.2213\\
ROUGE-3 & 0.1439 & 0.1457 & 0.1447\\
ROUGE-4 & 0.1058 & 0.1072 & 0.1064\\
ROUGE-L & 0.4091 & 0.4143 & 0.4114\\
% ROUGE-1 & 0.7150 & 0.3072 & 0.3965\\
% ROUGE-2 & 0.3836 & 0.1590 & 0.2088\\
% ROUGE-3 & 0.2530 & 0.1052 & 0.1383\\
% ROUGE-4 & 0.1853 & 0.0769 & 0.1013\\
% ROUGE-L & 0.6191 & 0.2660 & 0.3432\\
\end{tabular}
\caption{Title Driven based summaries at $depth=1$ with truncation.}
\label{table:dep1}
\end{table}

{\bf TITLE+BFS}. We can also produce summaries by traversing the full $TitleDrivenReduction$ tree and then truncating at 100 words. Those results are reported in Table~\ref{table:bfs}. Notice that this approach improves the F-1 score to $0.4751$. The main reason for this is due to the ordering of sentences at each level. Since, they are placed in document sentence order, this comes close to the strong $Lead$ method.

\begin{table}
% ../results/2018-0328-0402/rouge
\centering
\begin{tabular}{lccc}
Metric & Recall & Precision & F-Score\\
\hline
ROUGE-1 & 0.4687 & 0.4821 & 0.4751\\
ROUGE-2 & 0.2178 & 0.2239 & 0.2207\\
ROUGE-3 & 0.1424 & 0.1464 & 0.1443\\
ROUGE-4 & 0.1045 & 0.1075 & 0.1060\\
ROUGE-L & 0.4075 & 0.4192 & 0.4131\\
% ROUGE-1 & 0.4585 & 0.4867 & 0.4716\\
% ROUGE-2 & 0.2156 & 0.2286 & 0.2216\\
% ROUGE-3 & 0.1423 & 0.1509 & 0.1463\\
% ROUGE-4 & 0.1054 & 0.1118 & 0.1083\\
% ROUGE-L & 0.3985 & 0.4229 & 0.4098\\
\end{tabular}
\caption{Title Driven based summaries with breadth-first search and truncation.}
\label{table:bfs}
\end{table}

% \begin{table}[t]
% % ../results/2018-0328-0402/rouge
% \centering
% \begin{tabular}{lccc}
% Metric & Recall & Precision & F-Score\\
% \hline
% ROUGE-1 & 0.44685 & 0.48727 & 0.46340\\
% ROUGE-2 & 0.20690 & 0.22523 & 0.21453\\
% ROUGE-3 & 0.13562 & 0.14768 & 0.14067\\
% ROUGE-4 & 0.09954 & 0.10846 & 0.10329\\
% ROUGE-L & 0.38976 & 0.42489 & 0.40419\\
% ROUGE-W-1.2 & 0.15237 & 0.28014 & 0.19635\\
% \end{tabular}
% \caption{Title Driven based summaries at $depth=1$ with truncation.}
% \label{table:dep1}
% \end{table}

{\bf TITLE+FILTER}. As a final experiment, we use the title-driven process as a preprocessing step for our ILP formulations:
\begin{enumerate}
\item Use traditional techniques of stemming and stopword removal.
\item Apply Algorithm~\ref{algo::title} to reduce set of input sentences.
\item Apply $score\_ilp$ with $tfidf$ on reduced set of sentences.
\end{enumerate}

The best result found in the cubesearch was ROUGE1 F-1 score of $.4548$ with $r=-0.4$, $\alpha=1.2$ and $\beta=1.2$. This does not beat the breadth-first traversal of $TitleDrivenReduction$. 

\begin{figure}
    \centering
    \includegraphics[width=\columnwidth]{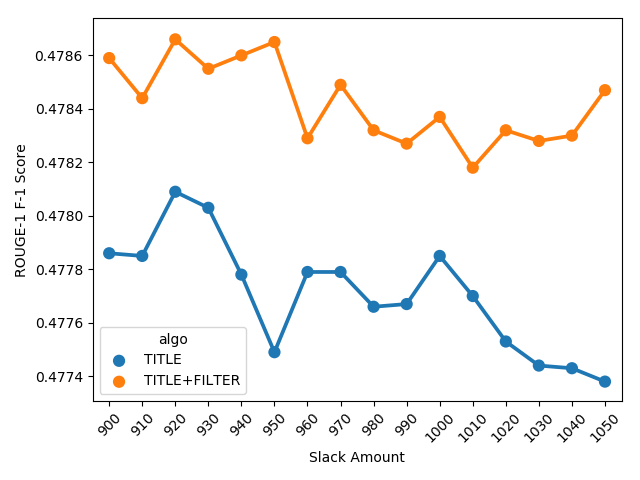}
    \caption{ROUGE-1 F-1 scores for different slack amounts using TITLE+FILTER ($TitleDrivenReduction$ then $score\_ilp$) and TITLE (only $score\_ilp$) on {\bf TITLE02} dataset.}
    \label{slack-t02}
\end{figure}

\begin{figure}
    \centering
    \includegraphics[width=\columnwidth]{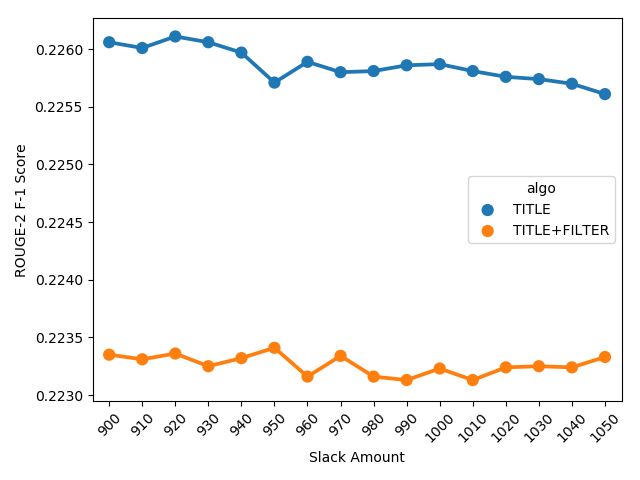}
    \caption{ROUGE-2 F-1 scores for different slack amounts using TITLE+FILTER ($TitleDrivenReduction$ then $score\_ilp$) and TITLE (only $score\_ilp$) on {\bf TITLE02} dataset.}
    \label{slack-t02-r2}
\end{figure}

{\bf TITLE+FILTER+SLACK} Because giving slack improves results, we also perform two experiments using the best parameter values for score ILP with {\bf TITLE+FILTER}. After step (3) we will then explore different amounts of slack, as we did for the Table~\ref{summ:truncate} experiment. For comparison, we do the same experiments without the $TitleDrivenReduction$. The experiments show that filtering and then using slack is the best performing unsupervised method (Figure~\ref{slack-t02}). Without the $TitleDrivenReduction$ using $score\_ilp$ with slack of $920$ gives a ROUGE-1 F-1 score $0.4782$. And when $TitleDrivenReduction$ is used as a filter, the unsupervised highest ROUGE-1 F-1 score of $0.4787$ (slack $= 920$) is achieved. We see a similar trend for ROUGE-L with scores of $0.4155$ and $0.4142$, with and without the filter, respectively (Appendix~\ref{appendix} Figure~\ref{slack-t02-rl}. However, for ROUGE-2 we see that applying a filter actually hurts the performance (Figure~\ref{slack-t02-r2}). We see the peak for ROUGE-2 is still at slack $=920$, but this changes when the filter is applied. Here the peak moves to slack $= 950$ and the score is $0.2235$. This leads to an intriguing new hypothesis that the more interesting bigrams in DUC02 dataset are in the sentences that do not directly overlap with the title. There could be deeper reasons behind this. We leave this investigation for future work. 
% ROUGE-2 and ROUGE-L also reach a maximum in the 900-1050 range. However, they plateau here. Without the filter, the scores are 

% \begin{table}
% \centering
% \begin{tabular}{lrr}
% Metric & Score ILP & Score ILP \& Title \\
% \hline
% ROUGE-1     & .4630 & .4628 \\
% ROUGE-2     & .2066 & .2066 \\
% ROUGE-3     & .1333 & .1333 \\
% ROUGE-4     & .0977 & .0977 \\
% ROUGE-LCS   & .3997 & .3998 \\
% ROUGE-W-1.2 & .1936 & .1937 \\
% \end{tabular}
% \caption{Need to verify these results. and add proper caption\textcolor{cyan}{I think we can remove this, if we have a cubesearch figure for title with ilp (i.e. $r$ $\alpha$ $\beta$.}}
% \end{table}

\subsection{Baseline 1 - Supervised Submodular}
While this system was designed for multi-document summarization, we adapted the system to be used for single document summarization. For the supervised submodular summarization algorithm, average ROUGE-1 F-1 score across five folds was $.5854$  on {\bf DUC02} dataset. For {\bf DUC01} it was $.5977$.
%Average summary length is 104.66. Min size is 79, Max size is 128 and Median is 104. Mean is 104.0. Histogram Figure~\ref{fig:subm-histo} 

%\textcolor{blue}{Daniel - check recall and precision and let's discuss more about this.}
%\textcolor{cyan}{
%In all cases recall is better by at least 12 points (thousandths) and as much as 24 points in the case of Fold 1.}
While the system was available for multi-document summarization, we tuned the system to be used for single document summarization. 
The submodular system was trained on the {\bf DUC02} using five-fold cross validation method. 
% The training took a total of 1 hr 27 mins on a machine with 64-bit Intel Xeon Processor with 128GB RAM.
The training took a total of 49 minutes and 29 seconds on a machine with 64-bit Intel Core i7 Processor with 16GB RAM.
The ROUGE evaluation results for the system have been presented in Table~\ref{tab:subm-title02}.

% added 2018-1219
\begin{table}
\centering\small
\begin{tabular}{llllll}
\hline
Metric&Fold 1&Fold 2&Fold 3&Fold 4&Fold 5 \\
\hline
R-1 & 0.8083 & 0.6863 & 0.5797 & 0.5252 & 0.3881 \\
R-2 & 0.6178 & 0.4950 & 0.3415 & 0.2653 & 0.1005 \\
R-3 & 0.4974 & 0.4100 & 0.2562 & 0.1753 & 0.0305 \\
R-4 & 0.3850 & 0.3434 & 0.1692 & 0.1354 & 0.0103 \\
R-L & 0.7979 & 0.6667 & 0.5314 & 0.4849 & 0.2886 \\
% R-W-1.2 & 0.4113 & 0.3256 & 0.2523 & 0.2250 & 0.1219 \\
\hline
\end{tabular}
\caption{F-1 scores, on {\bf TITLE02}, Submodular Five Folds. R-X above denotes ROUGE-X metric for X=1, 2, 3, 4, and L.}
\label{tab:subm-title02}
\end{table}

% added 2018-1219
% \begin{table}
% \centering\small
% \begin{tabular}{llllll}
% \hline
% Metric&Fold 1&Fold 2&Fold 3&Fold 4&Fold 5 \\
% \hline
% ROUGE-1 & 0.8083 & 0.6863 & 0.5797 & 0.5771 & 0.3861 \\
% ROUGE-2 & 0.6178 & 0.4950 & 0.3415 & 0.3317 & 0.1000 \\
% ROUGE-3 & 0.4974 & 0.4100 & 0.2562 & 0.2538 & 0.0303 \\
% ROUGE-4 & 0.3850 & 0.3434 & 0.1692 & 0.2154 & 0.0102 \\
% ROUGE-L & 0.7979 & 0.6667 & 0.5314 & 0.5274 & 0.2871 \\
% ROUGE-W-1.2 & 0.4113 & 0.3256 & 0.2523 & 0.2174 & 0.1215 \\
% \hline
% \end{tabular}
% \caption{F-1 scores, {\bf TITLE02}, Documents pre-processed with $TitleDrivenReduction$ and then Submodular Five Folds}
% \label{tab:duc02-subm-5fold}
% \end{table}

% added 2018-1219
% \begin{table}
% \centering\small
% \begin{tabular}{llllll}
% \hline
% Metric&Fold 1&Fold 2&Fold 3&Fold 4&Fold 5 \\
% \hline
% ROUGE-1 & 0.8085 & 0.6207 & 0.4000 & 0.5771 & 0.4221 \\
% ROUGE-2 & 0.6452 & 0.3781 & 0.2162 & 0.3317 & 0.1117 \\
% ROUGE-3 & 0.5326 & 0.3015 & 0.1644 & 0.2538 & 0.0308 \\
% ROUGE-4 & 0.4286 & 0.2538 & 0.1111 & 0.2154 & 0.0104 \\
% ROUGE-L & 0.7979 & 0.6010 & 0.3467 & 0.5274 & 0.3015 \\
% ROUGE-W-1.2 & 0.4057 & 0.2605 & 0.1687 & 0.2174 & 0.1286 \\
% \hline
% \end{tabular}
% \caption{F-1 scores, {\bf TITLE02}, pre-processing with $TitleDrivenReduction$ and $depth=1$, and then Submodular Five Folds}
% \label{tab:duc02-subm-5fold}
% \end{table}

We see that at the cost of more training time and the requirement of annotated data, we can get better results on summarization. We have included those results as a comparison to our unsupervised approach. For NewsSumm, there is no training time and producing summaries for the {\bf DUC02} dataset takes only 27 seconds. Since, the summaries can be quickly generated, it makes NewsSumm a strong candidate for preprocessing for the more expensive supervised methods.

Experiments were run to test the viability of preprocessing with NewsSumm. These experiments involve the $TitleDrivenReduction$ algorithm, the best algorithm in NewsSumm. Table~\ref{tab:submodular-all} shows how different parameters affect the results of submodular. It reports the average \mbox{ROUGE-1} F-1 Score, average summary length (in number of words), total training time, and extra time for running the algorithm. Running $TitleDrivenReduction$ produces smaller input files and allows submodular to decrease training time by 5 minutes and 44 seconds at the cost of only 55 seconds. All together, if preprocessing included as a part of training, it is a decrease of nearly 10\%. Not only this, the Average ROUGE-1 F-1 score improves as well. In addition, NewsSumm's $TitleDrivenReduction$ with $depth=1$ can dramatically decrease the training time. Again for a small preprocessing time of 53 seconds, there is a reduction of 58.3\%. There is a minor performance hit, but often the cost of training time can outweigh the gain in performance.

\begin{table*}
\centering\small    
	\begin{tabular}{lcrrr}
    \hline
    \textbf{EXP} & \textbf{5-fold Avg.} & \textbf{AVG SUMM LEN} & \textbf{Total Time} & \textbf{Additional Time (sec) } \\
    \hline
    Submodular Only & 0.5220 & 103.7 & 1881 & 0 \\
    % $TitleDrivenReduction$ & 0.6075 & 106.0 & 2625 & 55 \\
    $TitleDrivenReduction$ & 0.5260 & 103.3 & 1359 & 55 \\
    % $TitleDrivenReduction$ $depth=1$ & 0.5656 & 99.3 & 1238 & 53 \\
    $TitleDrivenReduction$ $depth=1$ & 0.4805 & 94.8 & 258 & 53 \\
    \hline
    \end{tabular}
    \caption{Supervised Submodular Results.}
    \label{tab:submodular-all}
\end{table*}

\subsection{Baseline 2 - PKUSUMSUM}
PKUSUMSUM offers several different unsupervised methods in their tool. We take a look at the summaries created by PKUSUMSUM.\footnote{We contacted author's for their system parameters, but have not received a response at the time of writing.}

\begin{table}
\centering\small
\begin{tabular}{llll}
\hline
\textbf{Alg.} & \textbf{ROUGE-1} & \textbf{ROUGE-2} & \textbf{ROUGE-L} \\ \hline
Centroid    & 0.4585 & 0.2135 & 0.4210 \\
Lead        & 0.4764 & 0.2248 & 0.4407 \\
LexPgRnk    & 0.3954 & 0.1326 & 0.3613 \\
Unsupervised & \multirow{2}{*}{0.4524} & \multirow{2}{*}{0.1805} & \multirow{2}{*}{0.4114}\\
\ \ Submodular &&&\\
TxtRnk      & 0.4296 & 0.1662 & 0.3871 \\ 
\hline
\end{tabular}
\caption{PKUSUMSUM ROUGE F-1 values on {\bf DUC02} dataset}
\label{tab:pku-duc02}
\end{table}

We see in Table~\ref{tab:pku-duc02} that in PKUSUMSUM the $Lead$ method is the best for news article summarization with Centroid second. 

\subsection{Summary Sizes}
Since the baselines, Supervised Submodular and PKUSUMSUM, are not implemented by us, we check the summary lengths for all top performing methods next. Table~\ref{tab:avg-sum-lth} reports the average summary lengths, along with standard deviations, on the DUC02 dataset. We see that the Lead method summaries of PKUSUMSUM are the longest on the average and Supervised Submodular summaries are second longest. NewsSumm summaries are slightly short of the ideal and Centroid has the shortest average summary length. Moreover, NewsSumm also has the smallest standard deviation. Thus, NewsSumm scores are achieved without the potential ``extra boost'' provided by longer summaries. 

For comparison, we report the results of the Baseline, which is also the first 100-word summary, from ~\citet{barrera:synsem11}. They reported a ROUGE-1 F-1 score of 0.4617. SynSem's highest ROUGE-1 F-1 score from~\citet{barrera:synsem11} is 0.4655 and S28's, the best system at DUC 02 competition, ROUGE-1 F-1 score  is 0.4673.  However, they do not report average summary length and standard deviations. Since the organizers of DUC02 truncated summaries to 100 words, we expect that S28 summaries should be in a narrow band around that number as well.
%We considered truncating the summaries of the other summarizers to 100 words, but decided against it, since it would not be fair to keep the shorter than 100 word su
\begin{table}
\centering\small
\begin{tabular}{lrr}
\hline
Summarizer & Average & Standard \\
    & Summary Length  & Deviation \\
\hline
Score\_ILP (NewsSumm) & 97.9 & 2.68 \\
Supervised Submodular &  104.0 & 6.98 \\
Lead (PKUSUMSUM) &   105.5 & 15.33 \\
Centroid (PKUSUMSUM) &  81.8 & 7.31 \\
Human (DUC02 dataset) &  101.1 & 4.20 \\
% R-W-1.2 & 0.4113 & 0.3256 & 0.2523 & 0.2250 & 0.1219 \\
\hline
\end{tabular}
\caption{Average Summary Length and Standard Deviation (in words) of Top Summarizers on DUC02 dataset.}
\label{tab:avg-sum-lth}
\end{table}
\subsection{Limit on Recall}
In~\citet{Verma:17}, we showed that there is a limit on ROUGE-1 F-
1-score of 0.907 for extractive summarizers on DUC02 dataset, and, as mentioned above, the limit is 0.9027 on TITLE02 dataset, so we see that the best extractive unsupervised summarizers are achieving about 52.7\% ($score\_ilp$) of that limit for DUC02 and 53.0\% (TITLE+FILTER+SLACK) for TITLE02 dataset. With supervised training we are able to improve to 64.5\% and 58.3\% on DUC02 and TITLE02, respectively.

\subsection{Qualitative Evaluation}\label{sec:qual}
A word cloud is a visual representation of a document that highlights the frequency counts of the words in the document. The relative differences in word frequencies can be represented through font size and/or colors. We use them here to give a qualitative analysis of the types of summaries being generated. We look specifically at one document set as defined by the organizers of DUC 2002. 

The document set we consider is $d069f$: a collection of 14 documents all considered relevant to a common topic. We compare the summaries generated by humans as well as four automatically generated summaries: lead method, PKUSUMSUM's centroid method, NewsSumm's best performing $score\_ilp$ method and Supervised Submodular with $TitleReduction$. For each word cloud, we create a document that combines all the documents. In addition, we remove all punctuation and change all letters to lowercase. This is then given as input to an online word cloud generator \footnote{\url{https://worditout.com/word-cloud/create}}. The top 100 most frequent words are selected to produce the final image. 

\begin{figure}
    \centering
    \begin{subfigure}{0.45\textwidth}
        \includegraphics[width=\columnwidth]{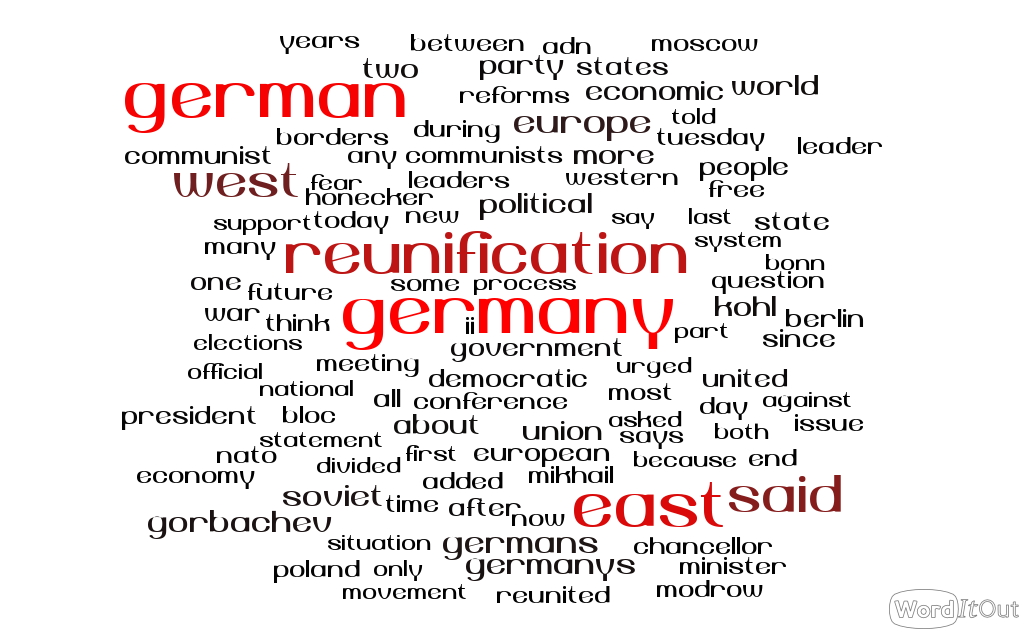}
        \caption{Original Documents of document set d069f.}
        \label{fig:original-d069f}
    \end{subfigure}
    \begin{subfigure}{0.45\textwidth}
        \includegraphics[width=\columnwidth]{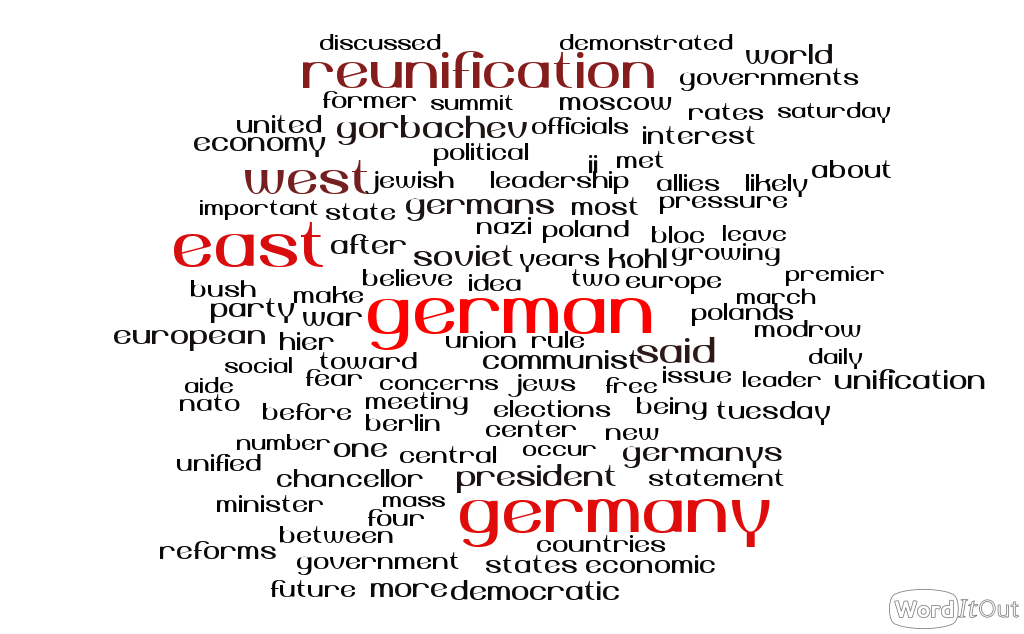}
        \caption{Human Summaries of document set d069f}
        \label{fig:humans-d069f}
    \end{subfigure}
\end{figure}

Figure~\ref{fig:original-d069f} represents the word cloud of document set $d069f$. Simply looking at the representation we can see clearly the content of the articles are on the reunification of Germany. The brighter red color and larger relevant size show words like ``germany'' and ``reunification'' are quite frequent in the document set. 

For comparison we can look at the word cloud of the human summaries in Figure~\ref{fig:humans-d069f}. Notice that ``german,'' ``germany,'' ``reunification,'' ``west,'' and ``east.'' are the most frequent in both human summaries and actual documents. To give a sense of what is not included in the summaries, we compare each summarizer with the original documents. We remove any words that are common to both the generated summaries and the original documents. 

\begin{figure*}
    \centering
    \begin{subfigure}{0.45\textwidth}
        \includegraphics[width=\textwidth]{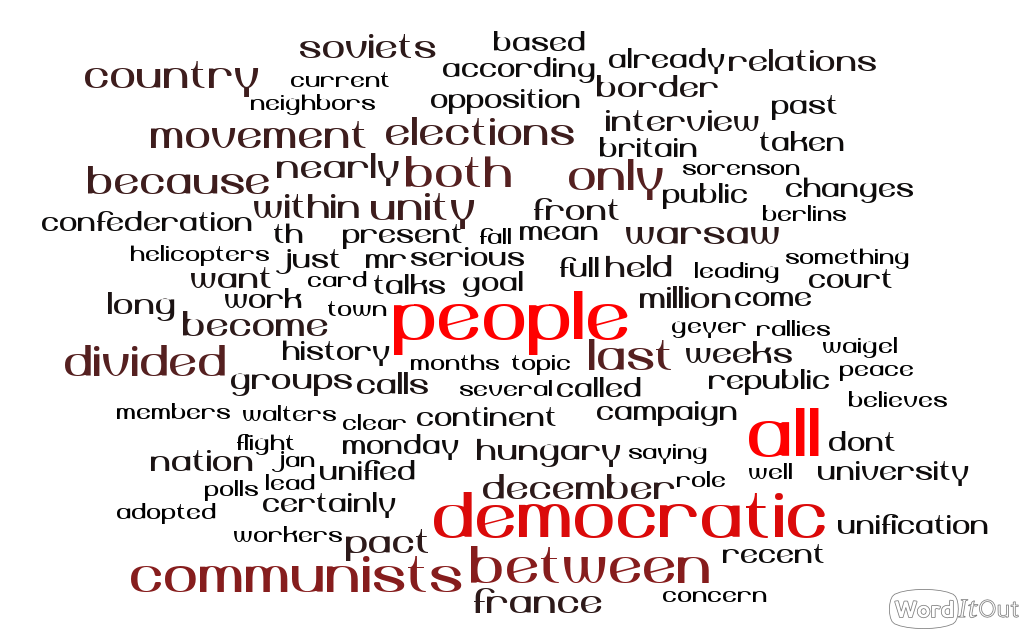}
        \caption{Lead method.}
    \end{subfigure}
    \begin{subfigure}{0.45\textwidth}
        \includegraphics[width=\textwidth]{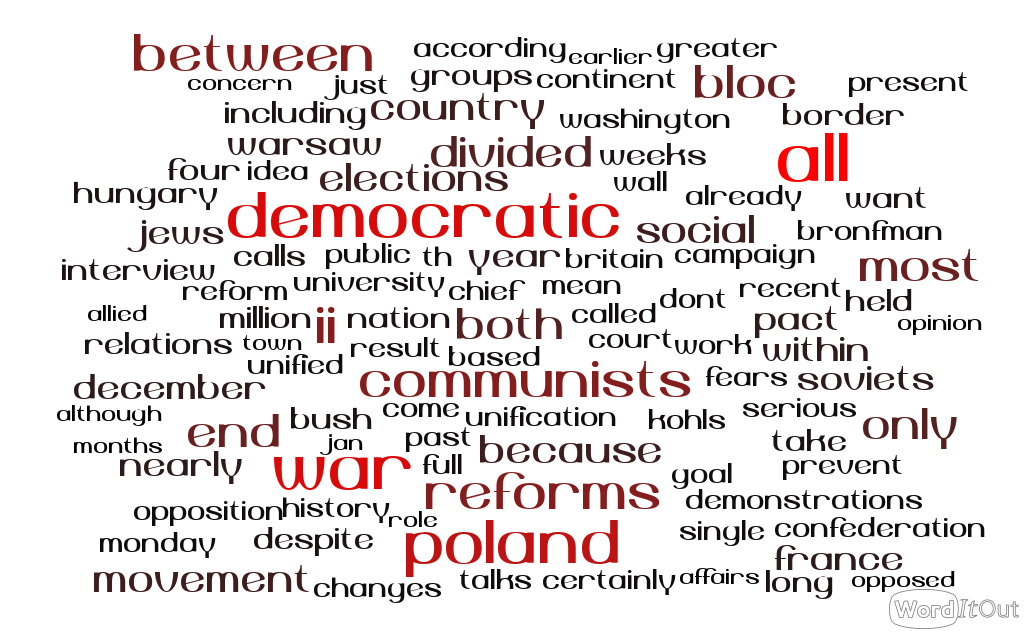}
        \caption{PKUSUMSUM Centroid.}
    \end{subfigure}
    \begin{subfigure}{0.45\textwidth}
        \includegraphics[width=\textwidth]{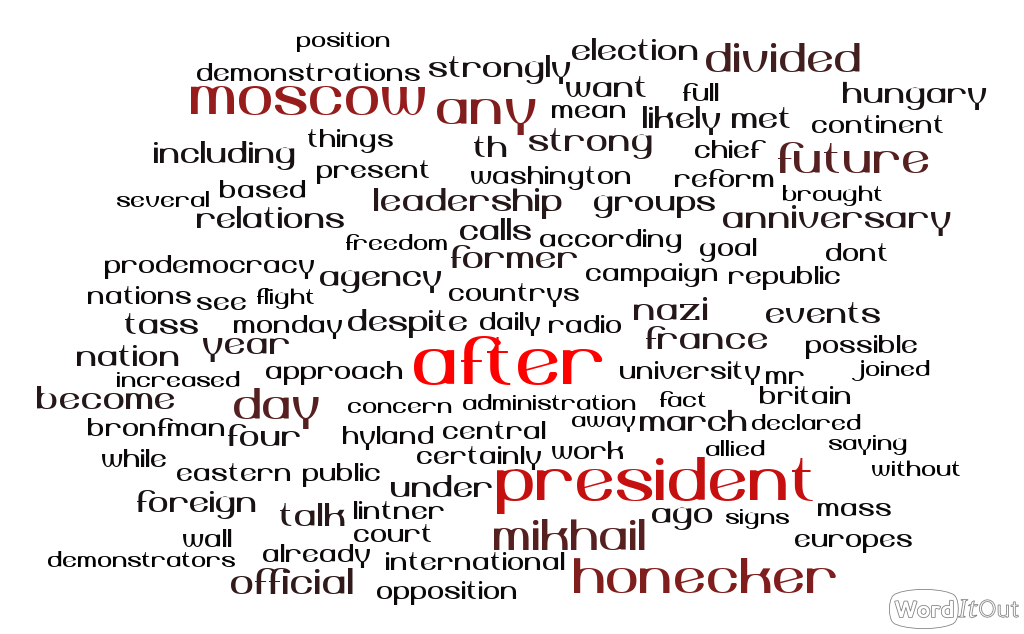}
        \caption{NewsSumm $score\_ilp$}
        \label{subf:newssumm}
    \end{subfigure}
    \begin{subfigure}{0.45\textwidth}
        \includegraphics[width=\textwidth]{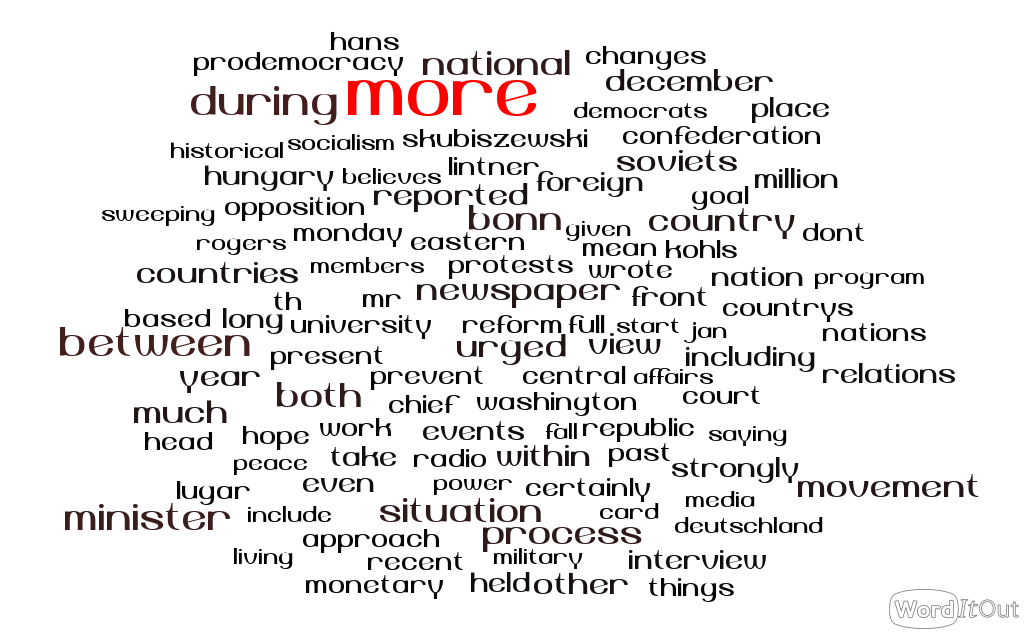}
        \caption{Supervised Submodular with Title Reduction.}
        \label{subf:subm}
    \end{subfigure}
    \begin{subfigure}{0.45\textwidth}
        \includegraphics[width=\textwidth]{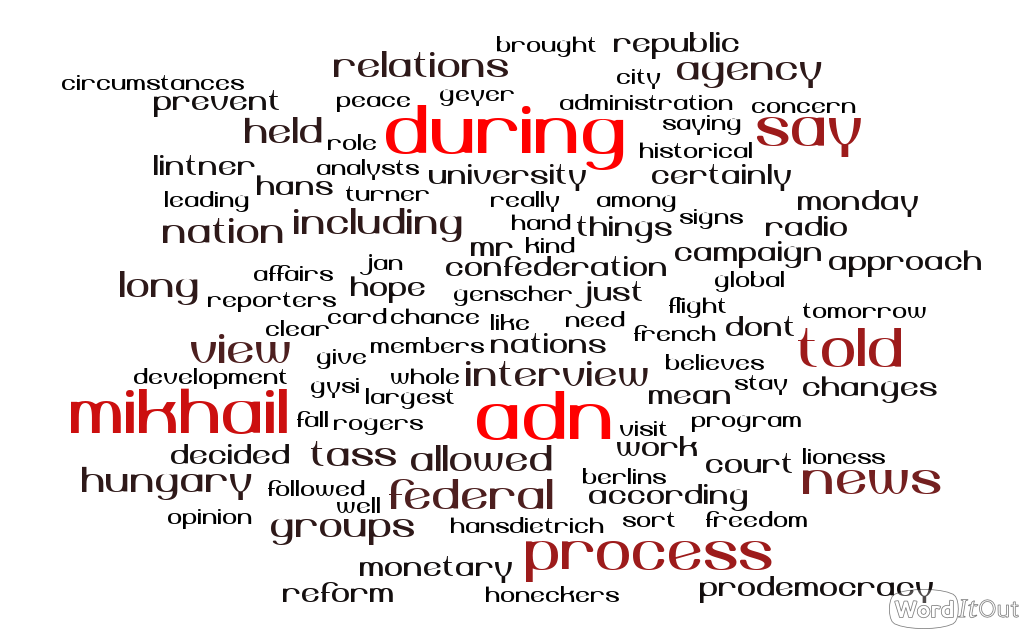}
        \caption{Human summaries. ({\em ``adn'' is not a typo but a news agency organization of Germany.})}
        \label{subf:humans}
    \end{subfigure}
    \caption{Each figure represents the remaining words of document set $d069f$ after the words of the generated summaries are removed.}
    \label{fig:doc-subtract}
\end{figure*}

We also look at the quality of the different summarizers by comparing unused words of the original documents. To be more specific, we compare words used in the summarizer and remove any occurrence of these words in the original documents. We then create a word cloud based on the frequencies of the remaining words. These can be seen in Figure~\ref{fig:doc-subtract}. For comparison we also included the original documents without the words used by human summaries (Figure~\ref{subf:humans}). 

The lead method and PKUSUMSUM's Centroid method are missing similar words (e.g. ``all'' and ``democratic''). Figure~\ref{subf:newssumm} shows NewsSumm misses words different from lead and PKUSUMSUM's Centroid methods. The content of NewsSumm summaries seem to be complementary to those of the lead method. And finally we see that the supervised submodular method captures a lot of the content in the original documents (evidenced by the lack of red-colored words). Also the remaining words in Figure~\ref{subf:subm} do not give a clear sense of what the documents are talking about. Like the supervised submodular method, we see that the human summaries are also including most of the indicative words.\footnote{The ``adn'' in wordcloud~\ref{subf:humans} is not a misspelt ``and,'' but the Agency ADN.} 

% \begin{figure}
%     \centering
%     \includegraphics[width=\columnwidth]{originals-small.png}
%     \caption{Original Documents of document set d069f.}
%     \label{fig:original-d069f}
% \end{figure}

% \begin{figure}
%     \centering
%     \includegraphics[width=\columnwidth]{humans-small.png}
%     \caption{Human Summaries of document set d069f}
%     \label{fig:humans-d069f}
% \end{figure}

\section{Conclusion}\label{sec:concl}
NewsSumm provides several avenues of experimentation for unsupervised document summarization. And although the algorithms do not out-perform supervised models, we show that we get competitive results in a trade-off for faster training times. Furthermore, this quick processing lends NewsSumm to be part of a larger pipeline. 

Deep learning models are currently dominating in machine learning tasks. Some researchers have moved on to more difficult tasks like abstract summarization or sentiment analysis. However, what is often overlooked is the training time. In this work we show that we can get good results at a fraction of the training time in the order of two magnitudes.

We have made use of third-party software (e.g. \citet{CPLEX}) and improved on this with novel algorithms. Continued research in unsupervised methods can be a step towards more robust solutions. NewsSumm is a step forward for quick implementation of unsupervised methods.

NewsSumm is a flexible tool for single-document summarization that provides the user several algorithms and options in a single, modular and easily-extensible framework. Even though its algorithms are not finely-tuned for any dataset, on ROUGE-2 and ROUGE-LCS metrics, it is already ahead of other finely-tuned systems on DUC datasets. We believe that researchers in summarization will benefit from having access to NewsSumm and it could also be deployed for customized filtering of the text flood confronting people. Extending NewsSumm for multi-document summarization is an interesting and useful task that we leave for the future. 

\bibliographystyle{cas-model2-names}
\bibliography{main}

\begin{thebibliography}{50}
\expandafter\ifx\csname natexlab\endcsname\relax\def\natexlab#1{#1}\fi
\providecommand{\url}[1]{\texttt{#1}}
\providecommand{\href}[2]{#2}
\providecommand{\path}[1]{#1}
\providecommand{\DOIprefix}{doi:}
\providecommand{\ArXivprefix}{arXiv:}
\providecommand{\URLprefix}{URL: }
\providecommand{\Pubmedprefix}{pmid:}
\providecommand{\doi}[1]{\href{http://dx.doi.org/#1}{\path{#1}}}
\providecommand{\Pubmed}[1]{\href{pmid:#1}{\path{#1}}}
\providecommand{\bibinfo}[2]{#2}
\ifx\xfnm\relax \def\xfnm[#1]{\unskip,\space#1}\fi
%Type = Article
\bibitem[{Baralis et~al.(2013)Baralis, Cagliero, Mahoto and
  Fiori}]{baralis2013graphsum}
\bibinfo{author}{Baralis, E.}, \bibinfo{author}{Cagliero, L.},
  \bibinfo{author}{Mahoto, N.}, \bibinfo{author}{Fiori, A.},
  \bibinfo{year}{2013}.
\newblock \bibinfo{title}{Graphsum: Discovering correlations among multiple
  terms for graph-based summarization}.
\newblock \bibinfo{journal}{Information Sciences} \bibinfo{volume}{249},
  \bibinfo{pages}{96--109}.
%Type = Inproceedings
\bibitem[{Barrera and Verma(2011)}]{barrera:synsem11}
\bibinfo{author}{Barrera, A.}, \bibinfo{author}{Verma, R.},
  \bibinfo{year}{2011}.
\newblock \bibinfo{title}{Automatic {E}xtractive {S}ingle-document
  {S}ummarization: {B}eating the {B}aselines with a {N}ew {A}pproach}, in:
  \bibinfo{booktitle}{Proceedings of the Symposium on Applied Computing},
  \bibinfo{organization}{ACM}.
%Type = Inproceedings
\bibitem[{Barrera and Verma(2012)}]{barrera2012combining}
\bibinfo{author}{Barrera, A.}, \bibinfo{author}{Verma, R.},
  \bibinfo{year}{2012}.
\newblock \bibinfo{title}{Combining syntax and semantics for automatic
  extractive single-document summarization}, in:
  \bibinfo{booktitle}{International Conference on Intelligent Text Processing
  and Computational Linguistics}, \bibinfo{organization}{Springer}. pp.
  \bibinfo{pages}{366--377}.
%Type = Inproceedings
\bibitem[{Carbonell and Goldstein(1998)}]{carbonell1998use}
\bibinfo{author}{Carbonell, J.}, \bibinfo{author}{Goldstein, J.},
  \bibinfo{year}{1998}.
\newblock \bibinfo{title}{The use of mmr, diversity-based reranking for
  reordering documents and producing summaries}, in:
  \bibinfo{booktitle}{Proceedings of the 21st annual international ACM SIGIR
  conference on Research and development in information retrieval},
  \bibinfo{organization}{ACM}. pp. \bibinfo{pages}{335--336}.
%Type = Inproceedings
\bibitem[{Cohan et~al.(2018)Cohan, Dernoncourt, Kim, Bui, Kim, Chang and
  Goharian}]{CohanDKBKCG18}
\bibinfo{author}{Cohan, A.}, \bibinfo{author}{Dernoncourt, F.},
  \bibinfo{author}{Kim, D.S.}, \bibinfo{author}{Bui, T.}, \bibinfo{author}{Kim,
  S.}, \bibinfo{author}{Chang, W.}, \bibinfo{author}{Goharian, N.},
  \bibinfo{year}{2018}.
\newblock \bibinfo{title}{A discourse-aware attention model for abstractive
  summarization of long documents}, in: \bibinfo{booktitle}{Proceedings of the
  2018 Conference of the North American Chapter of the Association for
  Computational Linguistics: Human Language Technologies, NAACL-HLT, New
  Orleans, Louisiana, USA, June 1-6, 2018, Volume 2 (Short Papers)}, pp.
  \bibinfo{pages}{615--621}.
%Type = Article
\bibitem[{Dong(2018)}]{dong18}
\bibinfo{author}{Dong, Y.}, \bibinfo{year}{2018}.
\newblock \bibinfo{title}{A survey on neural network-based summarization
  methods}.
\newblock \bibinfo{journal}{CoRR} \bibinfo{volume}{abs/1804.04589}.
\newblock \URLprefix \url{http://arxiv.org/abs/1804.04589}.
%Type = Article
\bibitem[{Erkan and Radev(2004)}]{erkan2004lexrank}
\bibinfo{author}{Erkan, G.}, \bibinfo{author}{Radev, D.R.},
  \bibinfo{year}{2004}.
\newblock \bibinfo{title}{Lexrank: Graph-based lexical centrality as salience
  in text summarization}.
\newblock \bibinfo{journal}{Journal of Artificial Intelligence Research}
  \bibinfo{volume}{22}, \bibinfo{pages}{457--479}.
%Type = Inproceedings
\bibitem[{Feigenblat et~al.(2017)Feigenblat, Boni, Roitman and
  Konopnicki}]{feigenblat2017summit}
\bibinfo{author}{Feigenblat, G.}, \bibinfo{author}{Boni, O.},
  \bibinfo{author}{Roitman, H.}, \bibinfo{author}{Konopnicki, D.},
  \bibinfo{year}{2017}.
\newblock \bibinfo{title}{Summit: A tool for extractive summarization,
  discovery and analysis}, in: \bibinfo{booktitle}{Proceedings of the 2017 ACM
  on Conference on Information and Knowledge Management},
  \bibinfo{organization}{ACM}. pp. \bibinfo{pages}{2459--2462}.
%Type = Article
\bibitem[{Ferreira et~al.(2013)Ferreira, de~Souza~Cabral, Lins, e~Silva,
  Freitas, Cavalcanti, Lima, Simske and Favaro}]{FerreiraCLSFCLSF13}
\bibinfo{author}{Ferreira, R.}, \bibinfo{author}{de~Souza~Cabral, L.},
  \bibinfo{author}{Lins, R.D.}, \bibinfo{author}{e~Silva, G.P.},
  \bibinfo{author}{Freitas, F.}, \bibinfo{author}{Cavalcanti, G.D.C.},
  \bibinfo{author}{Lima, R.}, \bibinfo{author}{Simske, S.J.},
  \bibinfo{author}{Favaro, L.}, \bibinfo{year}{2013}.
\newblock \bibinfo{title}{Assessing sentence scoring techniques for extractive
  text summarization}.
\newblock \bibinfo{journal}{Expert Syst. Appl.} \bibinfo{volume}{40},
  \bibinfo{pages}{5755--5764}.
\newblock \URLprefix \url{https://doi.org/10.1016/j.eswa.2013.04.023},
  \DOIprefix\doi{10.1016/j.eswa.2013.04.023}.
%Type = Article
\bibitem[{Gambhir and Gupta(2017)}]{gambhirG17}
\bibinfo{author}{Gambhir, M.}, \bibinfo{author}{Gupta, V.},
  \bibinfo{year}{2017}.
\newblock \bibinfo{title}{Recent automatic text summarization techniques: a
  survey}.
\newblock \bibinfo{journal}{Artif. Intell. Rev.} \bibinfo{volume}{47},
  \bibinfo{pages}{1--66}.
%Type = Inproceedings
\bibitem[{Garcia et~al.(2018)Garcia, Lima, Espinasse and Oliveira}]{garciaLE18}
\bibinfo{author}{Garcia, R.}, \bibinfo{author}{Lima, R.},
  \bibinfo{author}{Espinasse, B.}, \bibinfo{author}{Oliveira, H.},
  \bibinfo{year}{2018}.
\newblock \bibinfo{title}{Towards coherent single-document summarization: An
  integer linear programming-based approach}, in:
  \bibinfo{booktitle}{Proceedings of the 33rd Annual ACM Symposium on Applied
  Computing}, pp. \bibinfo{pages}{712--719}.
%Type = Article
\bibitem[{Garcia-Hernandez and Ledeneva(2009)}]{garciaherdandez:09}
\bibinfo{author}{Garcia-Hernandez, R.}, \bibinfo{author}{Ledeneva, Y.},
  \bibinfo{year}{2009}.
\newblock \bibinfo{title}{Word {S}equence {M}odels for {S}ingle {T}ext
  {S}ummarization}.
\newblock \bibinfo{journal}{Second Conf. on Advances in Computer-Human
  Interaction} , \bibinfo{pages}{44--48}.
%Type = Inproceedings
\bibitem[{Graham(2015)}]{Graham15}
\bibinfo{author}{Graham, Y.}, \bibinfo{year}{2015}.
\newblock \bibinfo{title}{Re-evaluating automatic summarization with {BLEU} and
  192 shades of {ROUGE}}, in: \bibinfo{booktitle}{Proceedings of the 2015
  Conference on Empirical Methods in Natural Language Processing, {EMNLP} 2015,
  Lisbon, Portugal, September 17-21, 2015}, pp. \bibinfo{pages}{128--137}.
\newblock \URLprefix \url{http://aclweb.org/anthology/D/D15/D15-1013.pdf}.
%Type = Inproceedings
\bibitem[{Gupta and Siddiqui(2012)}]{GuptaS12}
\bibinfo{author}{Gupta, V.K.}, \bibinfo{author}{Siddiqui, T.J.},
  \bibinfo{year}{2012}.
\newblock \bibinfo{title}{Multi-document summarization using sentence
  clustering}, in: \bibinfo{booktitle}{4th International Conference on
  Intelligent Human Computer Interaction, {IHCI} 2012, Kharagpur, India,
  December 27-29, 2012}, pp. \bibinfo{pages}{1--5}.
\newblock \URLprefix \url{https://doi.org/10.1109/IHCI.2012.6481826},
  \DOIprefix\doi{10.1109/IHCI.2012.6481826}.
%Type = Article
\bibitem[{IBM(2012)}]{CPLEX}
\bibinfo{author}{IBM}, \bibinfo{year}{2012}.
\newblock \bibinfo{title}{Ibm ilog cplex optimizer}.
\newblock \bibinfo{journal}{See www-01. ibm.
  com/software/integration/optimization/cplex-optimizer} .
%Type = Inproceedings
\bibitem[{Kleinberg et~al.(1999)Kleinberg, Kumar, Raghavan, Rajagopalan and
  Tomkins}]{kleinberg1999web}
\bibinfo{author}{Kleinberg, J.M.}, \bibinfo{author}{Kumar, R.},
  \bibinfo{author}{Raghavan, P.}, \bibinfo{author}{Rajagopalan, S.},
  \bibinfo{author}{Tomkins, A.S.}, \bibinfo{year}{1999}.
\newblock \bibinfo{title}{The web as a graph: measurements, models, and
  methods}, in: \bibinfo{booktitle}{International Computing and Combinatorics
  Conference}, \bibinfo{organization}{Springer}. pp. \bibinfo{pages}{1--17}.
%Type = Incollection
\bibitem[{Kumar et~al.(2013)Kumar, Srinathan and Varma}]{kumarKV13}
\bibinfo{author}{Kumar, N.}, \bibinfo{author}{Srinathan, K.},
  \bibinfo{author}{Varma, V.}, \bibinfo{year}{2013}.
\newblock \bibinfo{title}{A knowledge induced graph-theoretical model for
  extract and abstract single document summarization}, in:
  \bibinfo{booktitle}{Computational Linguistics and Intelligent Text
  Processing}. \bibinfo{publisher}{Springer}, pp. \bibinfo{pages}{408--423}.
%Type = Inproceedings
\bibitem[{Ledeneva et~al.(2011)Ledeneva, Garc{\'{\i}}a{-}Hern{\'{a}}ndez, Soto,
  Reyes and Gelbukh}]{LedenevaGSRG11}
\bibinfo{author}{Ledeneva, Y.},
  \bibinfo{author}{Garc{\'{\i}}a{-}Hern{\'{a}}ndez, R.A.},
  \bibinfo{author}{Soto, R.M.}, \bibinfo{author}{Reyes, R.C.},
  \bibinfo{author}{Gelbukh, A.F.}, \bibinfo{year}{2011}.
\newblock \bibinfo{title}{{EM} clustering algorithm for automatic text
  summarization}, in: \bibinfo{booktitle}{Advances in Artificial Intelligence -
  10th Mexican International Conference on Artificial Intelligence, {MICAI}
  2011, Puebla, Mexico, November 26 - December 4, 2011, Proceedings, Part {I}},
  pp. \bibinfo{pages}{305--315}.
\newblock \URLprefix \url{https://doi.org/10.1007/978-3-642-25324-9\_26},
  \DOIprefix\doi{10.1007/978-3-642-25324-9\_26}.
%Type = Article
\bibitem[{Li et~al.(2012)Li, Li and Li}]{li2012multi}
\bibinfo{author}{Li, J.}, \bibinfo{author}{Li, L.}, \bibinfo{author}{Li, T.},
  \bibinfo{year}{2012}.
\newblock \bibinfo{title}{Multi-document summarization via submodularity}.
\newblock \bibinfo{journal}{Applied Intelligence} \bibinfo{volume}{37},
  \bibinfo{pages}{420--430}.
%Type = Article
\bibitem[{Lin and Hovy(2003)}]{lin-hovy:03}
\bibinfo{author}{Lin, C.}, \bibinfo{author}{Hovy, E.}, \bibinfo{year}{2003}.
\newblock \bibinfo{title}{Automatic {E}valuation of {S}ummaries {U}sing n-gram
  {C}o-occurrence {S}tatistics}.
\newblock \bibinfo{journal}{HTL-NAACL} .
%Type = Inproceedings
\bibitem[{Lin(2004)}]{lin2004rouge}
\bibinfo{author}{Lin, C.Y.}, \bibinfo{year}{2004}.
\newblock \bibinfo{title}{Rouge: A package for automatic evaluation of
  summaries}, in: \bibinfo{booktitle}{Text summarization branches out:
  Proceedings of the ACL-04 workshop}, \bibinfo{organization}{Barcelona,
  Spain}.
%Type = Inproceedings
\bibitem[{Lin and Bilmes(2010)}]{lin2010multi}
\bibinfo{author}{Lin, H.}, \bibinfo{author}{Bilmes, J.}, \bibinfo{year}{2010}.
\newblock \bibinfo{title}{Multi-document summarization via budgeted
  maximization of submodular functions}, in: \bibinfo{booktitle}{Human Language
  Technologies: The 2010 Annual Conference of the North American Chapter of the
  Association for Computational Linguistics},
  \bibinfo{organization}{Association for Computational Linguistics}. pp.
  \bibinfo{pages}{912--920}.
%Type = Inproceedings
\bibitem[{Lin and Bilmes(2011)}]{lin2011class}
\bibinfo{author}{Lin, H.}, \bibinfo{author}{Bilmes, J.}, \bibinfo{year}{2011}.
\newblock \bibinfo{title}{A class of submodular functions for document
  summarization}, in: \bibinfo{booktitle}{Proceedings of the 49th Annual
  Meeting of the Association for Computational Linguistics: Human Language
  Technologies-Volume 1}, \bibinfo{organization}{Association for Computational
  Linguistics}. pp. \bibinfo{pages}{510--520}.
%Type = Inproceedings
\bibitem[{Liu et~al.(2018)Liu, Lu, Yang, Qu, Zhu and Li}]{LiuLYQZL18}
\bibinfo{author}{Liu, L.}, \bibinfo{author}{Lu, Y.}, \bibinfo{author}{Yang,
  M.}, \bibinfo{author}{Qu, Q.}, \bibinfo{author}{Zhu, J.},
  \bibinfo{author}{Li, H.}, \bibinfo{year}{2018}.
\newblock \bibinfo{title}{Generative adversarial network for abstractive text
  summarization}, in: \bibinfo{booktitle}{Proceedings of the Thirty-Second
  {AAAI} Conference on Artificial Intelligence, New Orleans, Louisiana, USA,
  February 2-7, 2018}.
%Type = Article
\bibitem[{Luhn(1958)}]{luhn1958automatic}
\bibinfo{author}{Luhn, H.P.}, \bibinfo{year}{1958}.
\newblock \bibinfo{title}{The automatic creation of literature abstracts}.
\newblock \bibinfo{journal}{IBM Journal of research and development}
  \bibinfo{volume}{2}, \bibinfo{pages}{159--165}.
%Type = Inproceedings
\bibitem[{Martins and Smith(2009)}]{martinsS09}
\bibinfo{author}{Martins, A.F.}, \bibinfo{author}{Smith, N.A.},
  \bibinfo{year}{2009}.
\newblock \bibinfo{title}{Summarization with a joint model for sentence
  extraction and compression}, in: \bibinfo{booktitle}{Proceedings of the
  Workshop on Integer Linear Programming for Natural Langauge Processing},
  \bibinfo{organization}{Association for Computational Linguistics}. pp.
  \bibinfo{pages}{1--9}.
%Type = Inproceedings
\bibitem[{McDonald(2007)}]{mcdonald2007study}
\bibinfo{author}{McDonald, R.}, \bibinfo{year}{2007}.
\newblock \bibinfo{title}{A study of global inference algorithms in
  multi-document summarization}, in: \bibinfo{booktitle}{European Conference on
  Information Retrieval}, \bibinfo{organization}{Springer}. pp.
  \bibinfo{pages}{557--564}.
%Type = Article
\bibitem[{Mendoza et~al.(2014)Mendoza, Bonilla, Noguera, Cobos and
  Le{\'o}n}]{mendoza2014extractive}
\bibinfo{author}{Mendoza, M.}, \bibinfo{author}{Bonilla, S.},
  \bibinfo{author}{Noguera, C.}, \bibinfo{author}{Cobos, C.},
  \bibinfo{author}{Le{\'o}n, E.}, \bibinfo{year}{2014}.
\newblock \bibinfo{title}{Extractive single-document summarization based on
  genetic operators and guided local search}.
\newblock \bibinfo{journal}{Expert Systems with Applications}
  \bibinfo{volume}{41}, \bibinfo{pages}{4158--4169}.
%Type = Inproceedings
\bibitem[{Mihalcea and Tarau(2004a)}]{mihalcea2004textrank}
\bibinfo{author}{Mihalcea, R.}, \bibinfo{author}{Tarau, P.},
  \bibinfo{year}{2004}a.
\newblock \bibinfo{title}{Textrank: Bringing order into texts},
  \bibinfo{organization}{Association for Computational Linguistics}.
%Type = Inproceedings
\bibitem[{Mihalcea and Tarau(2004b)}]{mihalcea:textrank04}
\bibinfo{author}{Mihalcea, R.}, \bibinfo{author}{Tarau, P.},
  \bibinfo{year}{2004}b.
\newblock \bibinfo{title}{Text{R}ank: {B}ringing {O}rder into {T}exts}, in:
  \bibinfo{booktitle}{Proceedings of the Conference on Empirical Methods in
  Natural Language Processing}, \bibinfo{organization}{(EMNLP, 2004)}.
%Type = Misc
\bibitem[{NIST(2014a)}]{duc2001info}
\bibinfo{author}{NIST}, \bibinfo{year}{2014}a.
\newblock \bibinfo{title}{{DUC 2001 Guidelines}}.
\newblock
  \bibinfo{howpublished}{\url{https://www-nlpir.nist.gov/projects/duc/guidelines/2001.html}}.
\newblock \bibinfo{note}{[Online; accessed 27-March-2018]}.
%Type = Misc
\bibitem[{NIST(2014b)}]{duc2002info}
\bibinfo{author}{NIST}, \bibinfo{year}{2014}b.
\newblock \bibinfo{title}{{DUC 2002 Guidelines}}.
\newblock
  \bibinfo{howpublished}{\url{https://www-nlpir.nist.gov/projects/duc/guidelines/2002.html}}.
\newblock \bibinfo{note}{[Online; accessed 27-March-2018]}.
%Type = Inproceedings
\bibitem[{Oliveira et~al.(2016)Oliveira, Lima, Lins, Freitas, Riss and
  Simske}]{oliveiraLL16}
\bibinfo{author}{Oliveira, H.}, \bibinfo{author}{Lima, R.},
  \bibinfo{author}{Lins, R.D.}, \bibinfo{author}{Freitas, F.},
  \bibinfo{author}{Riss, M.}, \bibinfo{author}{Simske, S.J.},
  \bibinfo{year}{2016}.
\newblock \bibinfo{title}{Assessing concept weighting in integer linear
  programming based single-document summarization}, in:
  \bibinfo{booktitle}{Proceedings of the 2016 ACM Symposium on Document
  Engineering}, pp. \bibinfo{pages}{205--208}.
%Type = Inproceedings
\bibitem[{Owczarzak et~al.(2012)Owczarzak, Conroy, Dang and
  Nenkova}]{OwczarzakCDN12}
\bibinfo{author}{Owczarzak, K.}, \bibinfo{author}{Conroy, J.M.},
  \bibinfo{author}{Dang, H.T.}, \bibinfo{author}{Nenkova, A.},
  \bibinfo{year}{2012}.
\newblock \bibinfo{title}{An assessment of the accuracy of automatic evaluation
  in summarization}, in: \bibinfo{booktitle}{Proceedings of Workshop on
  Evaluation Metrics and System Comparison for Automatic
  Summarization@NACCL-HLT 2012, Montr{\`{e}}al, Canada, June 2012, 2012}, pp.
  \bibinfo{pages}{1--9}.
\newblock \URLprefix \url{https://aclanthology.info/papers/W12-2601/w12-2601}.
%Type = Techreport
\bibitem[{Page et~al.(1999)Page, Brin, Motwani and Winograd}]{page1999pagerank}
\bibinfo{author}{Page, L.}, \bibinfo{author}{Brin, S.},
  \bibinfo{author}{Motwani, R.}, \bibinfo{author}{Winograd, T.},
  \bibinfo{year}{1999}.
\newblock \bibinfo{title}{The PageRank citation ranking: Bringing order to the
  web.}
\newblock \bibinfo{type}{Technical Report}. Stanford InfoLab.
%Type = Inproceedings
\bibitem[{Polsley et~al.(2016)Polsley, Jhunjhunwala and
  Huang}]{polsley2016casesummarizer}
\bibinfo{author}{Polsley, S.}, \bibinfo{author}{Jhunjhunwala, P.},
  \bibinfo{author}{Huang, R.}, \bibinfo{year}{2016}.
\newblock \bibinfo{title}{Casesummarizer: A system for automated summarization
  of legal texts}, in: \bibinfo{booktitle}{Proceedings of COLING 2016, the 26th
  International Conference on Computational Linguistics: System
  Demonstrations}, pp. \bibinfo{pages}{258--262}.
%Type = Inproceedings
\bibitem[{Rachabathuni(2017)}]{rachabathuni2017survey}
\bibinfo{author}{Rachabathuni, P.K.}, \bibinfo{year}{2017}.
\newblock \bibinfo{title}{A survey on abstractive summarization techniques},
  in: \bibinfo{booktitle}{Inventive Computing and Informatics (ICICI),
  International Conference on}, \bibinfo{organization}{IEEE}. pp.
  \bibinfo{pages}{762--765}.
%Type = Article
\bibitem[{Radev et~al.(2004)Radev, Jing, Sty and Tam}]{RadevJST04}
\bibinfo{author}{Radev, D.R.}, \bibinfo{author}{Jing, H.},
  \bibinfo{author}{Sty, M.}, \bibinfo{author}{Tam, D.}, \bibinfo{year}{2004}.
\newblock \bibinfo{title}{Centroid-based summarization of multiple documents}.
\newblock \bibinfo{journal}{Inf. Process. Manage.} \bibinfo{volume}{40},
  \bibinfo{pages}{919--938}.
\newblock \URLprefix \url{https://doi.org/10.1016/j.ipm.2003.10.006},
  \DOIprefix\doi{10.1016/j.ipm.2003.10.006}.
%Type = Inproceedings
\bibitem[{See et~al.(2017)See, Liu and Manning}]{see2017get}
\bibinfo{author}{See, A.}, \bibinfo{author}{Liu, P.J.},
  \bibinfo{author}{Manning, C.D.}, \bibinfo{year}{2017}.
\newblock \bibinfo{title}{Get to the point: Summarization with
  pointer-generator networks}, in: \bibinfo{booktitle}{Proceedings of the 55th
  Annual Meeting of the Association for Computational Linguistics (Volume 1:
  Long Papers)}, pp. \bibinfo{pages}{1073--1083}.
%Type = Inproceedings
\bibitem[{Sipos and Joachims(2013)}]{sipos2013generating}
\bibinfo{author}{Sipos, R.}, \bibinfo{author}{Joachims, T.},
  \bibinfo{year}{2013}.
\newblock \bibinfo{title}{Generating comparative summaries from reviews}, in:
  \bibinfo{booktitle}{Proceedings of the 22nd ACM international conference on
  Conference on information \& knowledge management},
  \bibinfo{organization}{ACM}. pp. \bibinfo{pages}{1853--1856}.
%Type = Inproceedings
\bibitem[{Sipos et~al.(2012a)Sipos, Shivaswamy and Joachims}]{sipos2012large}
\bibinfo{author}{Sipos, R.}, \bibinfo{author}{Shivaswamy, P.},
  \bibinfo{author}{Joachims, T.}, \bibinfo{year}{2012}a.
\newblock \bibinfo{title}{Large-margin learning of submodular summarization
  models}, in: \bibinfo{booktitle}{Proceedings of the 13th Conference of the
  European Chapter of the Association for Computational Linguistics},
  \bibinfo{organization}{Association for Computational Linguistics}. pp.
  \bibinfo{pages}{224--233}.
%Type = Inproceedings
\bibitem[{Sipos et~al.(2012b)Sipos, Swaminathan, Shivaswamy and
  Joachims}]{sipos2012temporal}
\bibinfo{author}{Sipos, R.}, \bibinfo{author}{Swaminathan, A.},
  \bibinfo{author}{Shivaswamy, P.}, \bibinfo{author}{Joachims, T.},
  \bibinfo{year}{2012}b.
\newblock \bibinfo{title}{Temporal corpus summarization using submodular word
  coverage}, in: \bibinfo{booktitle}{Proceedings of the 21st ACM international
  conference on Information and knowledge management},
  \bibinfo{organization}{ACM}. pp. \bibinfo{pages}{754--763}.
%Type = Article
\bibitem[{Sparck~Jones(1972)}]{sparck1972statistical}
\bibinfo{author}{Sparck~Jones, K.}, \bibinfo{year}{1972}.
\newblock \bibinfo{title}{A statistical interpretation of term specificity and
  its application in retrieval}.
\newblock \bibinfo{journal}{Journal of documentation} \bibinfo{volume}{28},
  \bibinfo{pages}{11--21}.
%Type = Inproceedings
\bibitem[{Svore et~al.(2007)Svore, Vanderwende and Burges}]{svore2007enhancing}
\bibinfo{author}{Svore, K.M.}, \bibinfo{author}{Vanderwende, L.},
  \bibinfo{author}{Burges, C.J.}, \bibinfo{year}{2007}.
\newblock \bibinfo{title}{Enhancing single-document summarization by combining
  ranknet and third-party sources.}, in: \bibinfo{booktitle}{Emnlp-conll}, pp.
  \bibinfo{pages}{448--457}.
%Type = Article
\bibitem[{Verma and Lee(2017)}]{Verma:17}
\bibinfo{author}{Verma, R.M.}, \bibinfo{author}{Lee, D.}, \bibinfo{year}{2017}.
\newblock \bibinfo{title}{Extractive summarization: Limits, compression,
  generalized model and heuristics}.
\newblock \bibinfo{journal}{Computaci{\'{o}}n y Sistemas} \bibinfo{volume}{21}.
\newblock \URLprefix
  \url{http://www.cys.cic.ipn.mx/ojs/index.php/CyS/article/view/2855}.
%Type = Book
\bibitem[{Williams et~al.(2009)}]{williams2009logic}
\bibinfo{author}{Williams, H.P.}, et~al., \bibinfo{year}{2009}.
\newblock \bibinfo{title}{Logic and integer programming}. volume
  \bibinfo{volume}{130}.
\newblock \bibinfo{publisher}{Springer}.
%Type = Inproceedings
\bibitem[{Xu et~al.(2017)Xu, Lau, Baldwin and Cohn}]{xu2017decoupling}
\bibinfo{author}{Xu, Y.}, \bibinfo{author}{Lau, J.H.},
  \bibinfo{author}{Baldwin, T.}, \bibinfo{author}{Cohn, T.},
  \bibinfo{year}{2017}.
\newblock \bibinfo{title}{Decoupling encoder and decoder networks for
  abstractive document summarization}, in: \bibinfo{booktitle}{Proceedings of
  the MultiLing 2017 Workshop on Summarization and Summary Evaluation Across
  Source Types and Genres}, pp. \bibinfo{pages}{7--11}.
%Type = Article
\bibitem[{Yang et~al.(2014)Yang, Cai, Zhang and Shi}]{yang2014enhancing}
\bibinfo{author}{Yang, L.}, \bibinfo{author}{Cai, X.}, \bibinfo{author}{Zhang,
  Y.}, \bibinfo{author}{Shi, P.}, \bibinfo{year}{2014}.
\newblock \bibinfo{title}{Enhancing sentence-level clustering with
  ranking-based clustering framework for theme-based summarization}.
\newblock \bibinfo{journal}{Information sciences} \bibinfo{volume}{260},
  \bibinfo{pages}{37--50}.
%Type = Inproceedings
\bibitem[{Zhang et~al.(2016)Zhang, Wang and Wan}]{zhang2016pkusumsum}
\bibinfo{author}{Zhang, J.}, \bibinfo{author}{Wang, T.}, \bibinfo{author}{Wan,
  X.}, \bibinfo{year}{2016}.
\newblock \bibinfo{title}{Pkusumsum: a java platform for multilingual document
  summarization}, in: \bibinfo{booktitle}{Proceedings of COLING 2016, the 26th
  International Conference on Computational Linguistics: System
  Demonstrations}, pp. \bibinfo{pages}{287--291}.
%Type = Inproceedings
\bibitem[{Zhang and Li(2009)}]{zhangL09}
\bibinfo{author}{Zhang, P.y.}, \bibinfo{author}{Li, C.h.},
  \bibinfo{year}{2009}.
\newblock \bibinfo{title}{Automatic text summarization based on sentences
  clustering and extraction}, in: \bibinfo{booktitle}{Computer Science and
  Information Technology, 2009. ICCSIT 2009. 2nd IEEE International Conference
  on}, \bibinfo{organization}{IEEE}. pp. \bibinfo{pages}{167--170}.

\end{thebibliography}

%%%%%%%%%%%%%%%%%%%%%
%%%%%%% OTHER %%%%%%%
%%%%%%%%%%%%%%%%%%%%%

% \clearpage
\appendix
\section{Bad or no headlines}
\label{app:bad-headline}
Some of the original XMLs used the ``{\tt <HEAD>}'' field to signal the article was a follow-up to a previous article (e.g. ``\emph{With BC-APN-Oscars}''). Others seemed like internal communications between editors and journalists (e.g. ``\emph{Eds: To update with Bush attending service, adds new graf after 4th
previous, The Iowa}''). The remaining did not include any title. The following articles contained no valid headline:
\begin{itemize}
\item AP880720-0262 (no title), 
\item AP900328-0128 (no title), 
\item AP880712-0250 (no title),
\item AP880328-0206 (internal messaging),
\item AP890420-0176 (internal messaging), 
\item AP891116-0191 (internal messaging), 
\item and AP890119-0221 (internal messaging)
\end{itemize}

\section{No Overlapping Title}
\label{no-overlap}
Following are a list of documents where the title did not overlap with any body text. We have included the title of those documents as well for reference:
\begin{itemize}
\item AP900210-0106: ``Thousands Demonstrate for Unification''
\item LA080290-0037: ``HOW THE CONFLICT DEVELOPED''
\item LA102089-0177: ``THE BAY AREA QUAKE; WHAT NEXT?; PONDERING THE LESSONS, HEALING THE SCARS''
\item LA111289-0035: ``ROGER SIMON: A STRATEGY THAT BEARS REPEATING''
\end{itemize}

\section{Supplementary Figures}
\label{appendix}
These figures provide ROUGE-2 F-1 scores for comparison.
\FloatBarrier
\begin{figure}
	\centering
    \includegraphics[height=5cm,keepaspectratio]{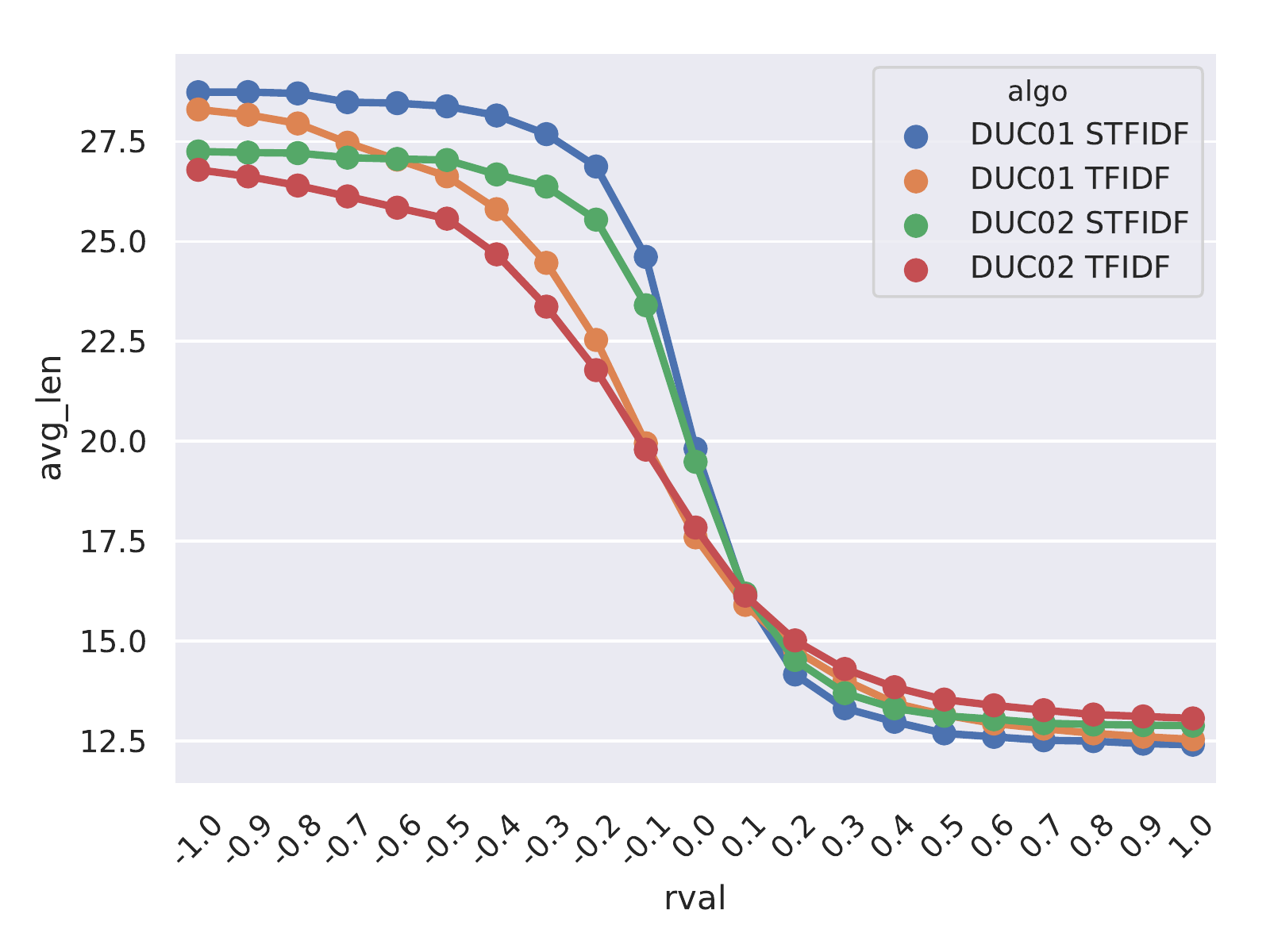}
    \caption{Plot of average length of summaries across different $r$-values.}
    \label{fig:tf2stf-avg-len}
\end{figure}
\begin{figure}
    \centering
    \includegraphics[height=5cm,keepaspectratio]{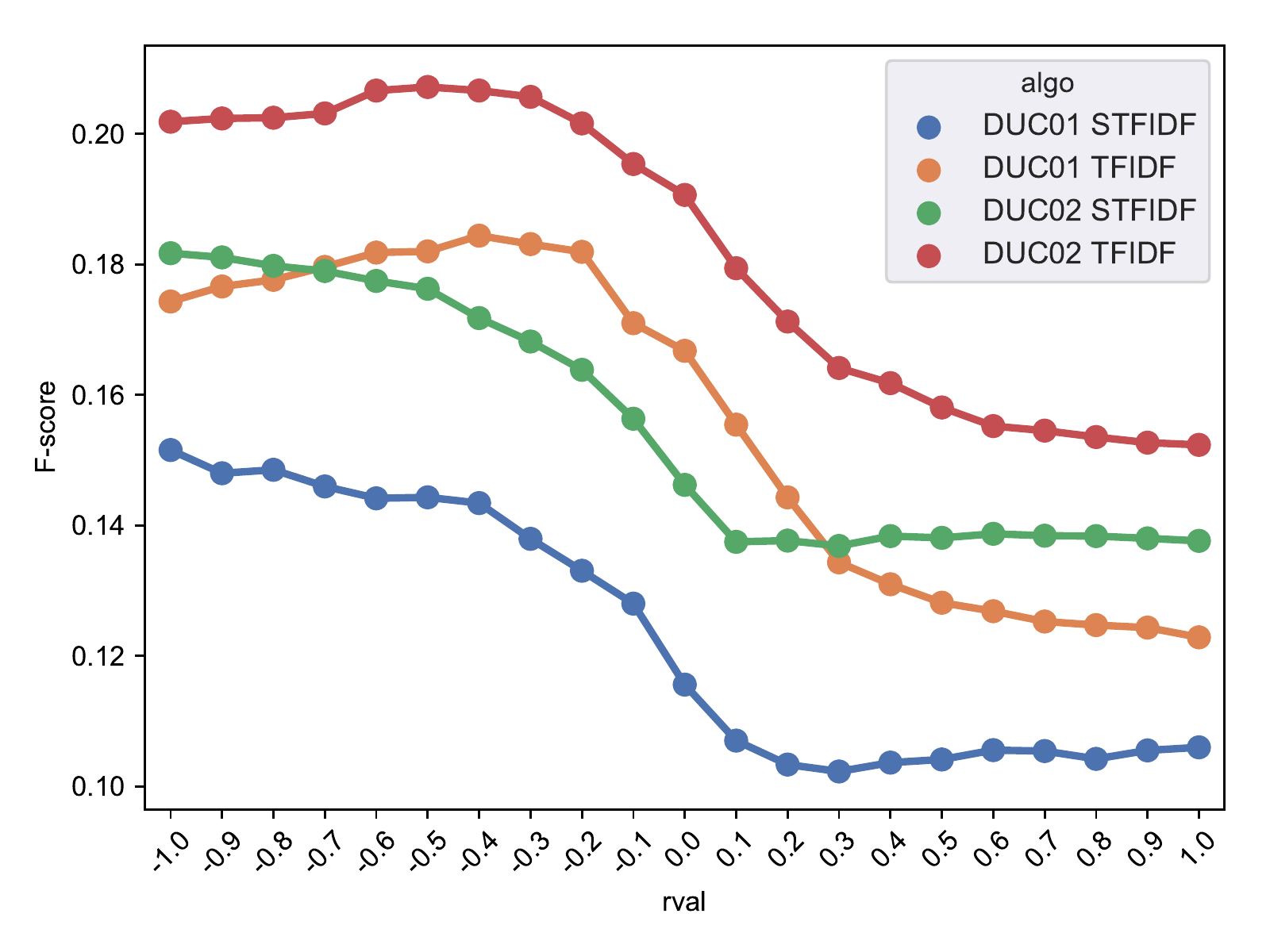}
    \caption{ROUGE-2 F-scores on {\bf DUC02} and {\bf DUC01} at different $r$-values.}
\end{figure}
\begin{figure}
\centering
    \includegraphics[height=5cm,keepaspectratio]{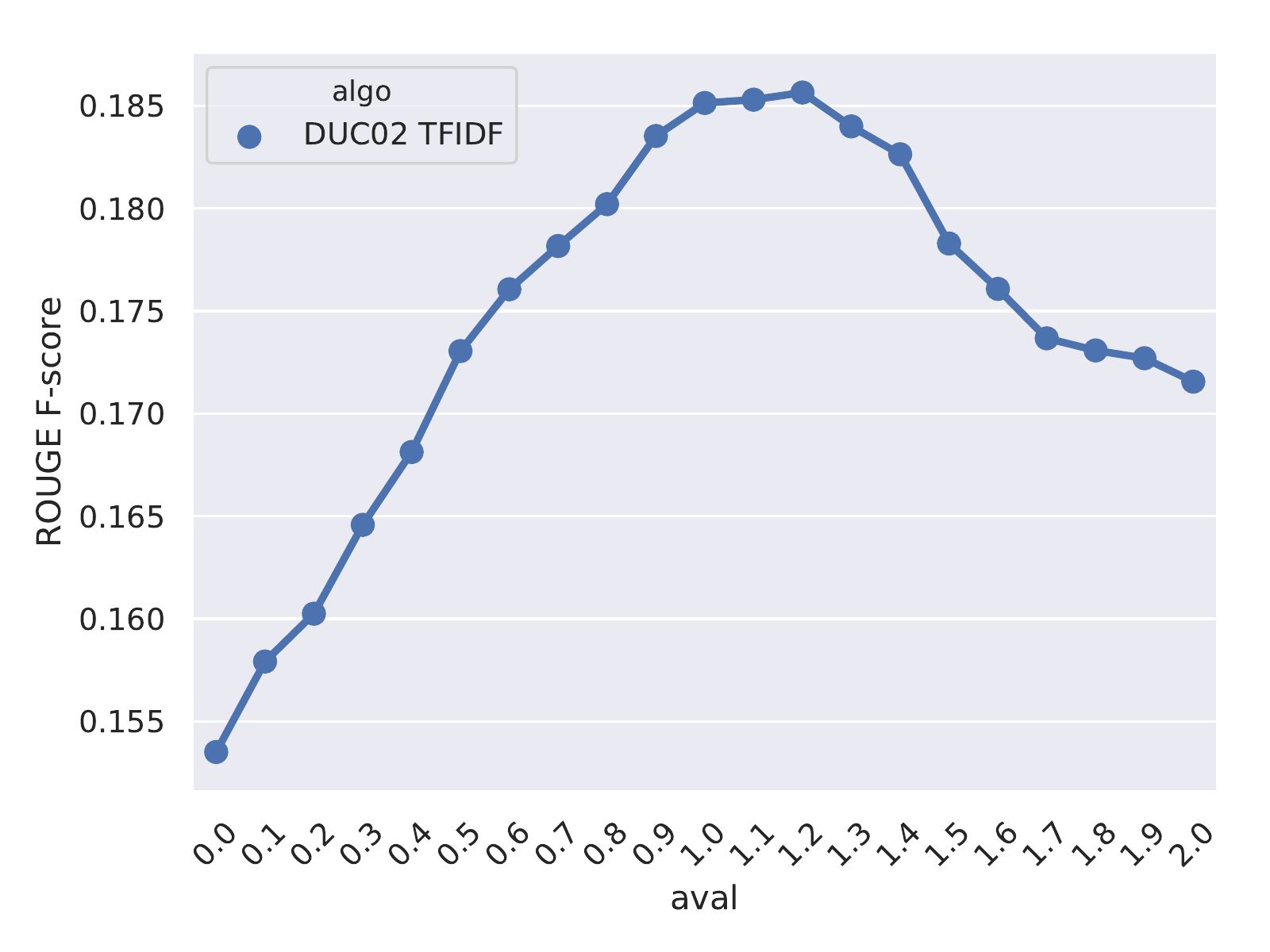}
    \caption{Effect of $\alpha$ ($\beta =1$) on {\bf DUC02} ROUGE-2 F-1 scores using $tfidf$ and $score\_ilp$.}
\end{figure}
\begin{figure}
    \centering
    \includegraphics[height=5cm,keepaspectratio]{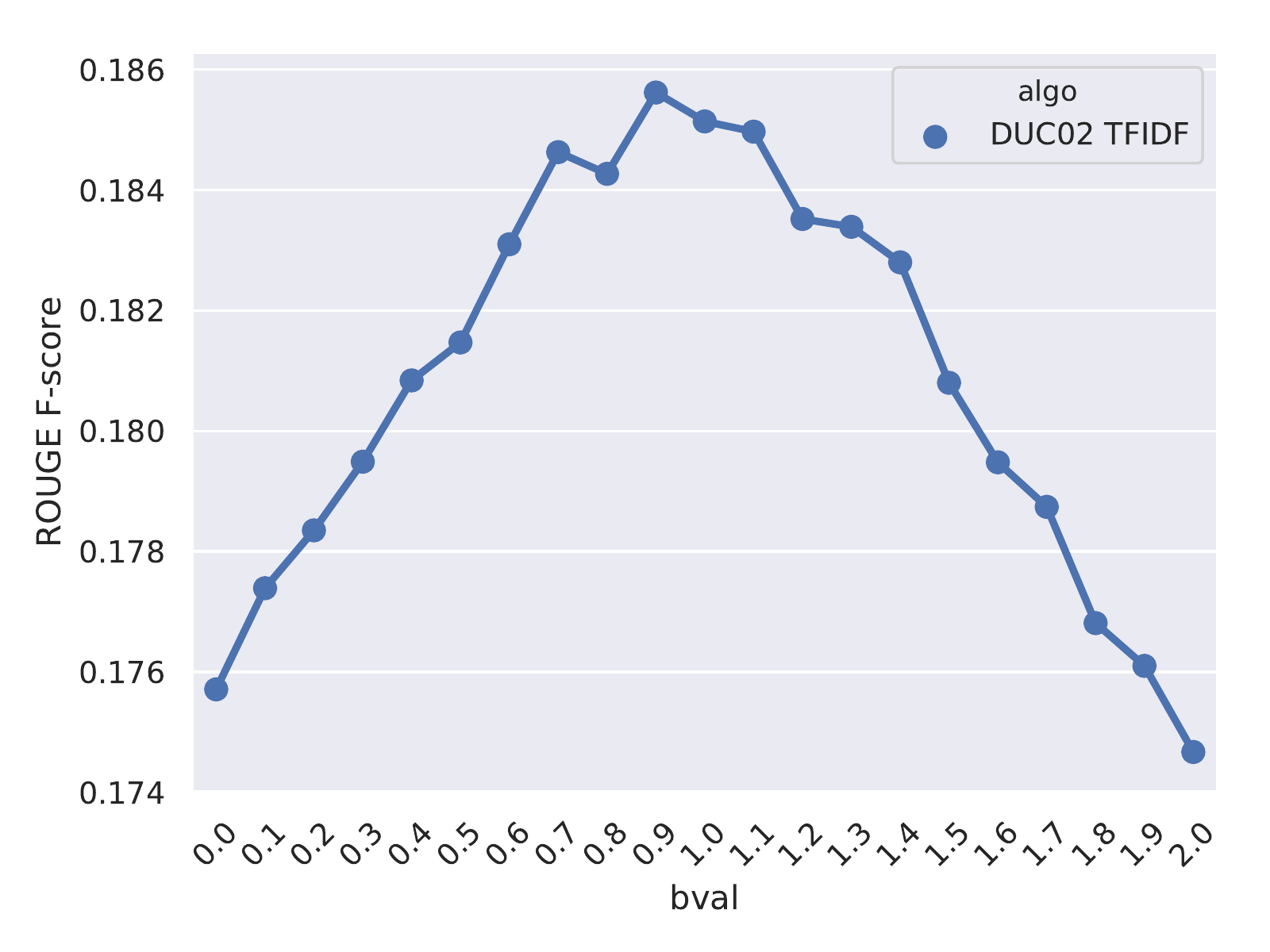}
    \caption{Effect of $\beta$ ($\alpha = 1$) on {\bf DUC02} ROUGE-2 F-1 scores using $tfidf$ and $score$.}
\end{figure}
\begin{figure}
    \centering
    \includegraphics[height=5cm,keepaspectratio]{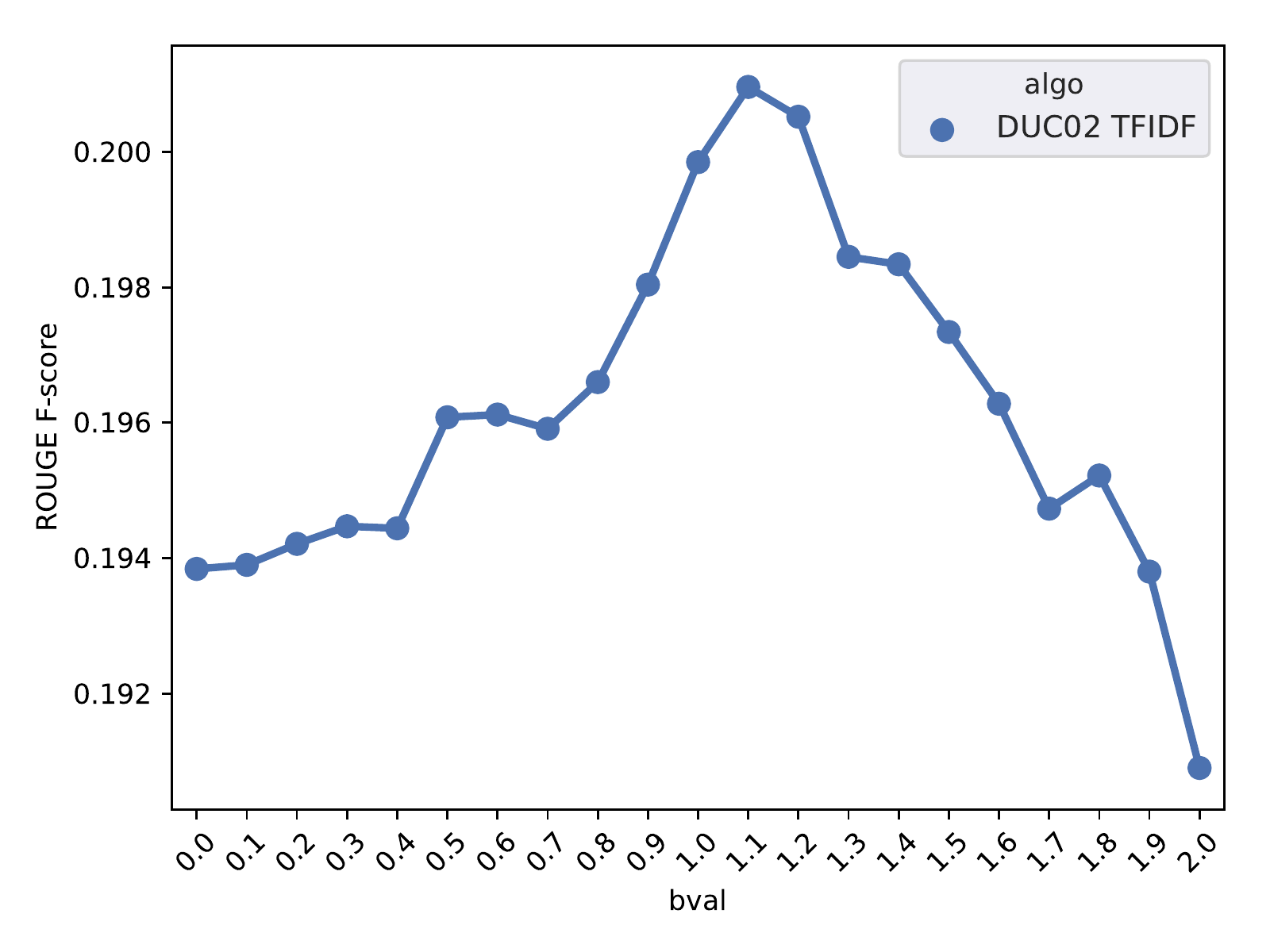}
    \caption{ROUGE-2 F-scores on {\bf DUC02} for varying $\beta$ with $r=-0.4$ and $\alpha=1.1$}
\end{figure}
\begin{figure}
    \centering
    \includegraphics[height=5cm,keepaspectratio]{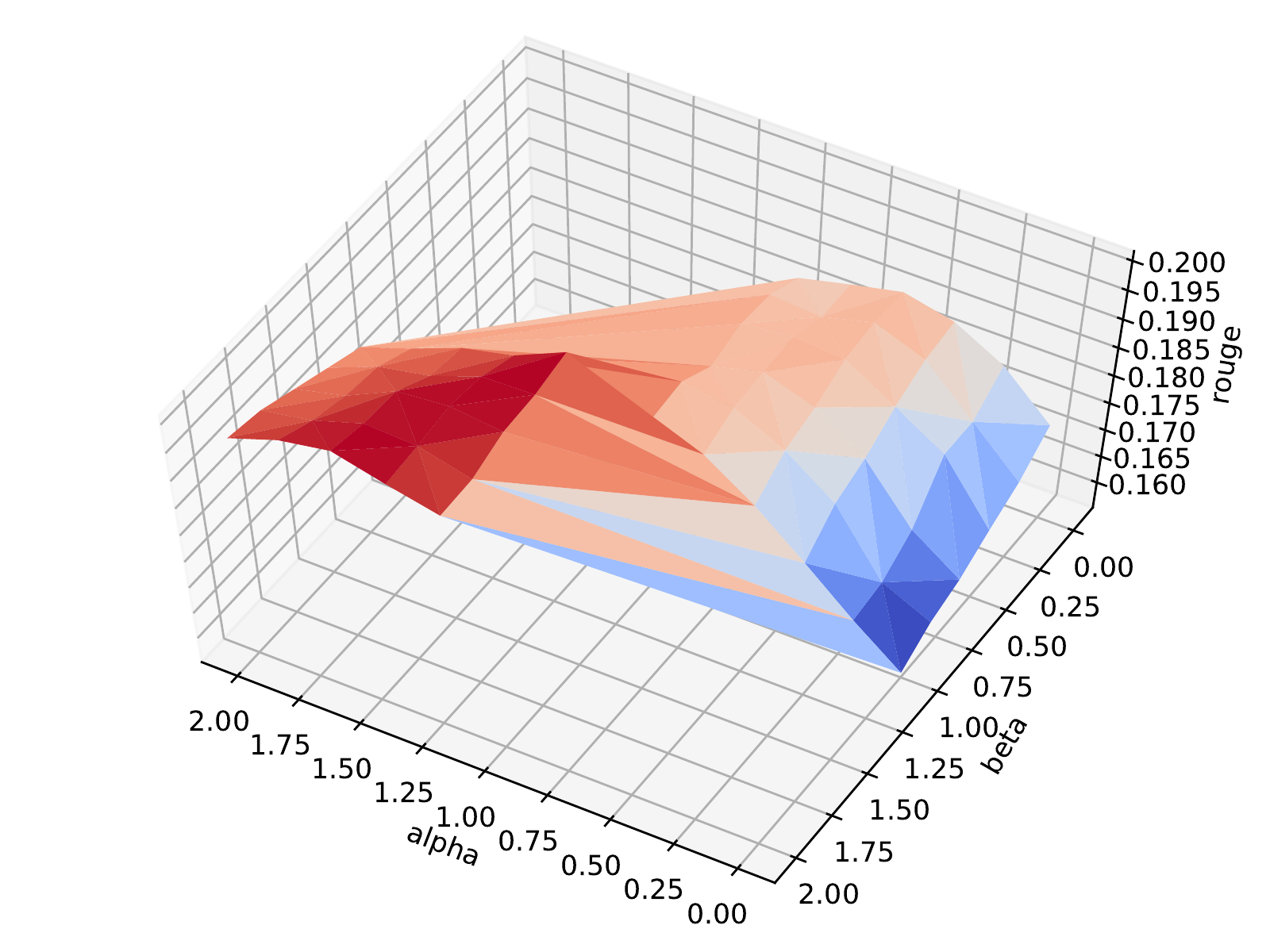}
    \caption{ROUGE-2 F-score on {\bf DUC02} varying $\alpha$ and $\beta$ and $r$-value fixed at $-0.4$.}
    \label{cubesearch-duc02-r2}
\end{figure}
\begin{figure}
    \centering
    \includegraphics[height=5cm,keepaspectratio]{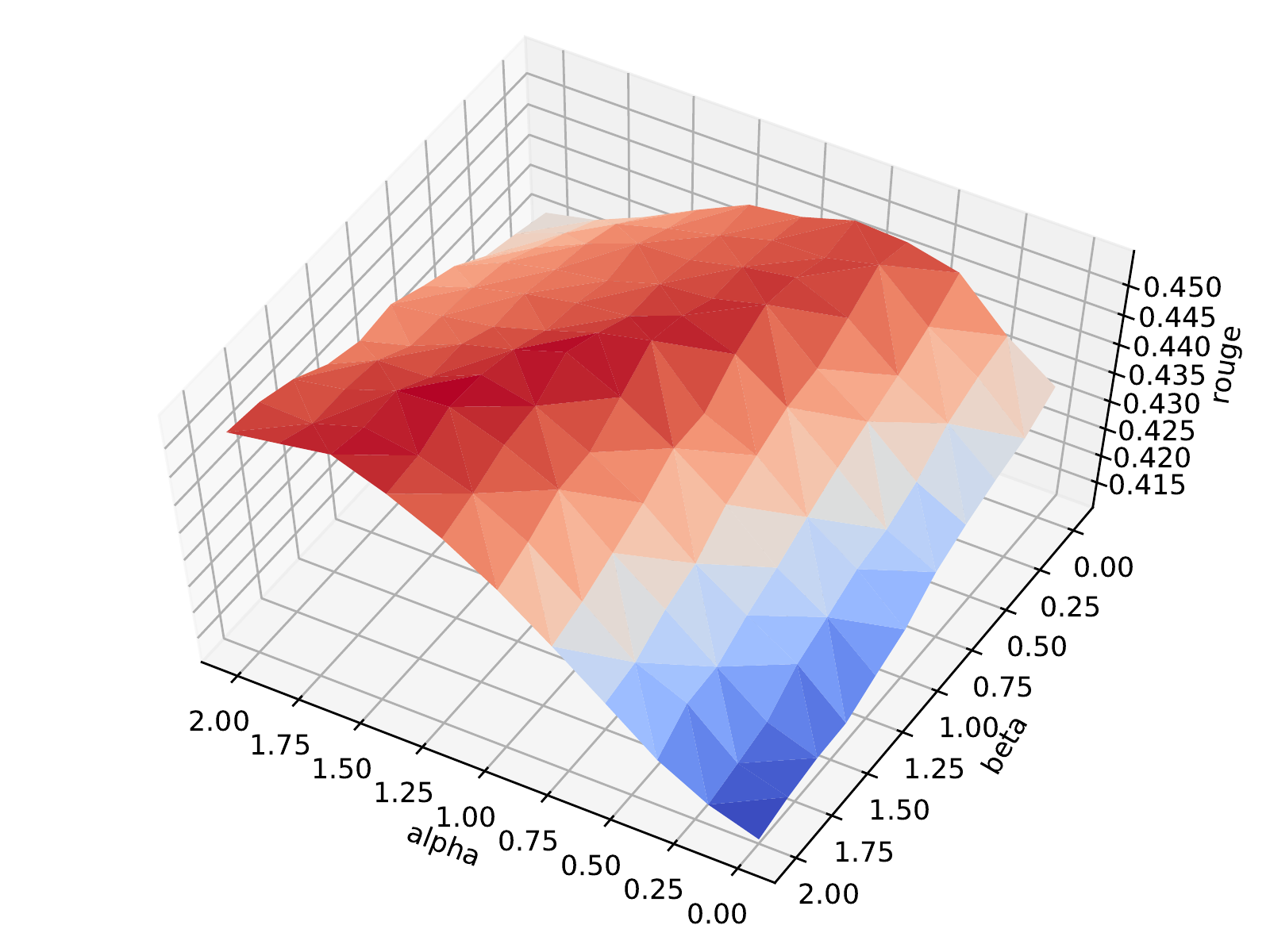}
    \caption{ROUGE-1 F-score on {\bf TITLE02} varying $\alpha$ and $\beta$ and $r$-value fixed at $-0.4$.}
    \label{cubesearch-title02}
\end{figure}
\begin{figure}
    \centering
    \includegraphics[height=5cm,keepaspectratio]{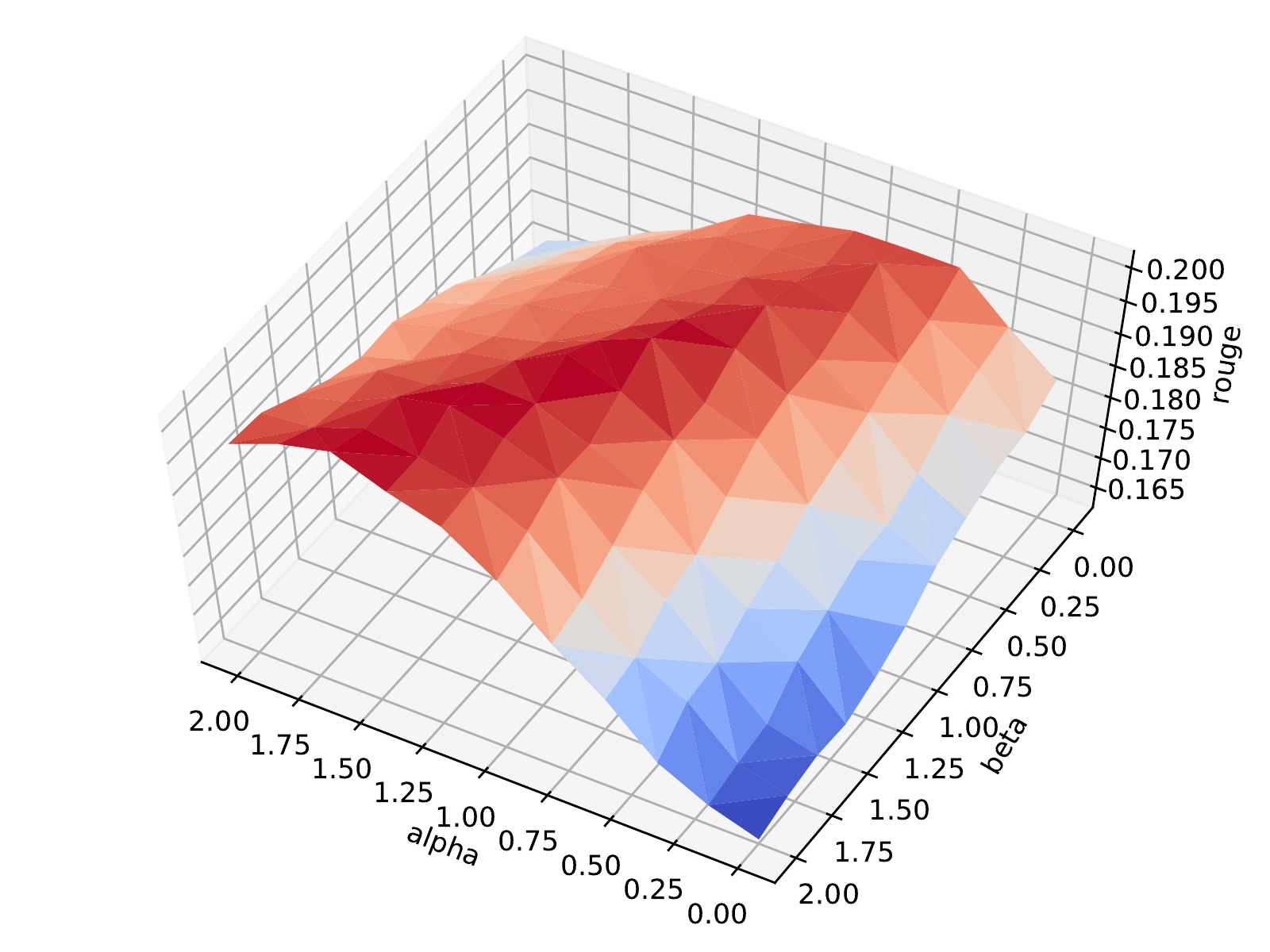}
    \caption{ROUGE-2 F-score on {\bf TITLE02} varying $\alpha$ and $\beta$ and $r$-value fixed at $-0.4$.}
    \label{cubesearch-title02-r2}
\end{figure}
\begin{figure}
    \centering
    \includegraphics[height=5cm,keepaspectratio]{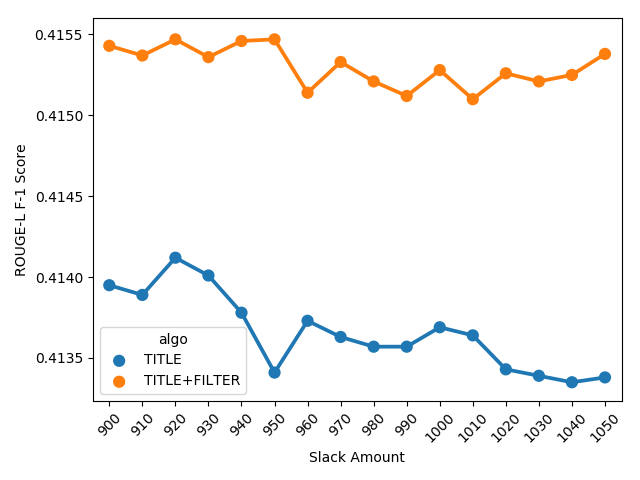}
    \caption{ROUGE-L F-1 scores for different slack amounts using TITLE+FILTER ($TitleDrivenReduction$ then $score\_ilp$) and TITLE (only $score\_ilp$) on {\bf TITLE02} dataset.}
    \label{slack-t02-rl}
\end{figure}
\FloatBarrier

\end{document}